\documentclass[manuscript, authorversion]{acmart}
\usepackage{enumitem}
\setlist[itemize]{leftmargin=*}
\usepackage{xcolor}
\usepackage{changepage}
\usepackage{array}
\usepackage{colortbl}

\usepackage{algorithm}
\usepackage{algpseudocode}
\algnewcommand\algorithmicforeach{\textbf{for each}}
\algdef{S}[FOR]{ForEach}[1]{\algorithmicforeach\ #1\ \algorithmicdo}

\newenvironment{examplecallout}
  {\begin{adjustwidth}{1em}{1em}%
   \color{black!75}%
   \itshape
   \noindent
   \hangindent=1.8em
   \hangafter=1}
  {\par\end{adjustwidth}}

\newcommand{\sys}{{ValiAssist}{}}
\usepackage{enumitem}
\setlist[itemize]{leftmargin=*}

\AtBeginDocument{%
  }

\setcopyright{acmlicensed}
\copyrightyear{2018}
\acmYear{2018}
\acmDOI{XXXXXXX.XXXXXXX}
\acmConference[Conference acronym 'XX]{Make sure to enter the correct
  conference title from your rights confirmation email}{June 03--05,
  2018}{Woodstock, NY}
\acmISBN{978-1-4503-XXXX-X/2018/06}

\begin{document}

\title{\sys{}: Scaling LLM-generated Programming Feedback Validation by Demonstration}
\renewcommand{\shortauthors}{Trovato et al.}
\author{Xiaohang Tang}
\affiliation{%
  \institution{Virginia Tech}
  \city{Blacksburg}
  \state{Virginia}
  \country{USA}
}
\email{xiaohangtang@vt.edu}

\author{Rehema Abulikemu}
\affiliation{%
  \institution{Virginia Tech}
  \city{Blacksburg}
  \state{Virginia}
  \country{USA}
}
\email{rexime@vt.edu}

\author{Peng Gao}
\affiliation{%
  \institution{Virginia Tech}
  \city{Blacksburg}
  \state{Virginia}
  \country{USA}
}
\email{penggao@vt.edu}

\author{Yan Chen}
\affiliation{%
  \institution{Virginia Tech}
  \city{Blacksburg}
  \state{Virginia}
  \country{USA}
}
\email{ych@vt.edu}
\begin{abstract}
Generative AI can create programming feedback at scale, but validating its alignment with instructors’ pedagogical goals and strategies remains challenging. Instructors may struggle to fully specify validation criteria upfront, as these often emerge through inspecting concrete examples and revising misaligned feedback. We present \sys{}, a validation-by-demonstration system that infers reusable validation logic from instructors’ revisions of AI-generated feedback and allows instructors to inspect and refine that logic as it is applied across other student submissions. In a comparative study with 16 instructors, ValiAssist helped instructors create and refine validation criteria, producing feedback that better aligned with their pedagogical intent (91\% vs. 57\%) and more than doubling the accuracy of correcting problematic criteria (77\% vs. 32\%). Our findings suggest that validation by demonstration can help instructors turn judgments they know how to make on individual cases into explicit, reusable logic, while progressively clarifying where those judgments do and do not generalize.


\end{abstract}


\begin{teaserfigure}
  \includegraphics[width=\textwidth]{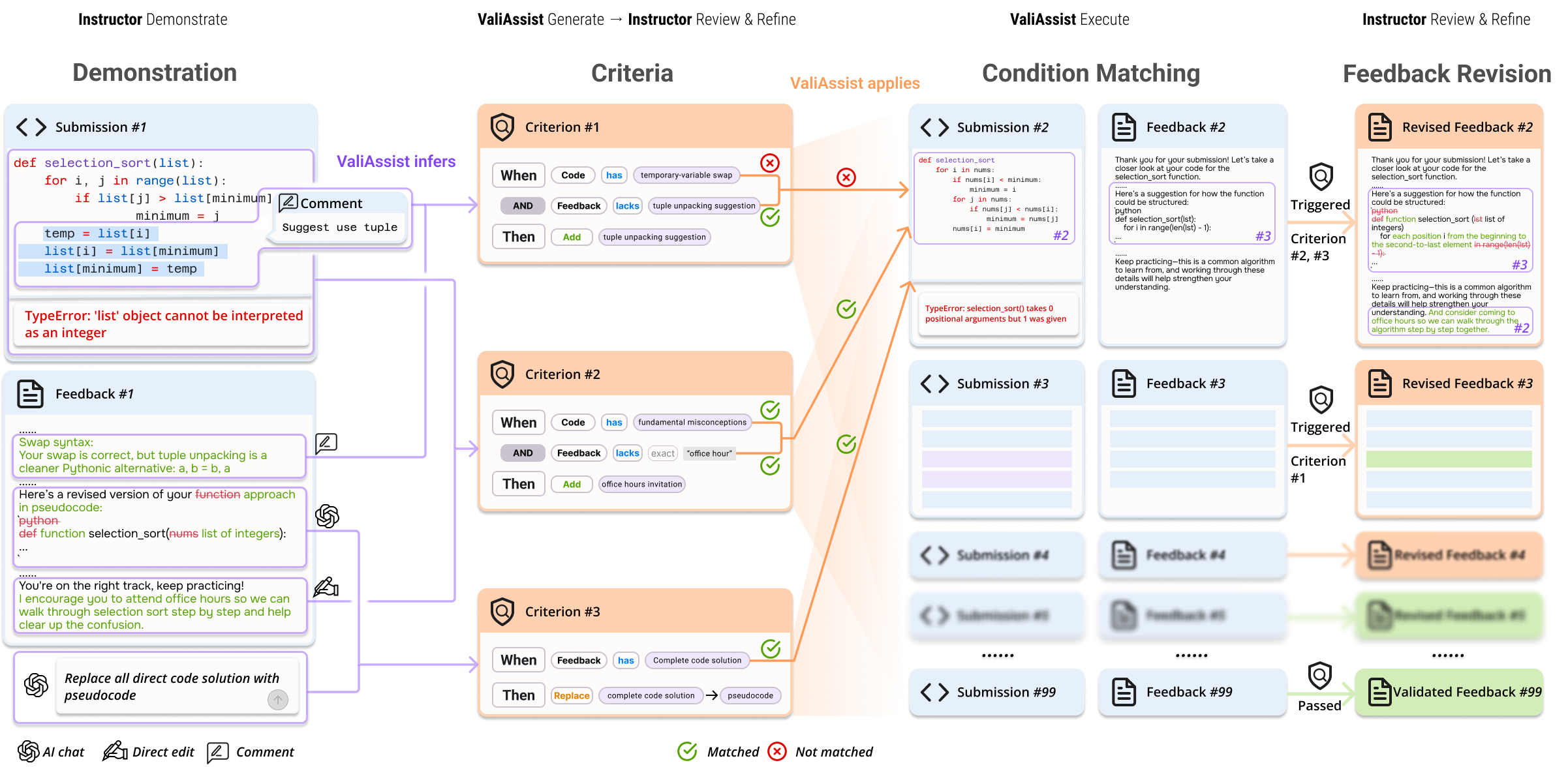}
  \caption{The validation-by-demonstration workflow of \sys. The instructor starts by revising LLM-generated feedback in a sample code-feedback pair, then \sys{} suggests validation criteria by inferring the instructor's intention according to their interactions (e.g., comments, AI chat, and direct edit) and context information (e.g., original and revised feedback, code submission, and error message). After verifying and tuning the criteria, the instructor can apply the criteria to all other code-feedback pairs to scale the validation. \sys{} first performs condition matching on the code-feedback pairs and takes actions based on the matched criteria.}
  \label{fig:teaser}
\end{teaserfigure}
\maketitle
\section{Introduction}

Providing timely, individualized feedback to students is valuable but difficult to sustain at scale. LLMs offer a way to generate programming feedback across many student submissions ~\cite{nguyen2024using, kazemitabaar2024codeaid, nguyen2024comparing}, but such feedback may be incorrect, conflict with course policies or learning objectives, or provide guidance that instructors consider pedagogically inappropriate ~\cite{kalai2025language, kazemitabaar2024codeaid}. Because these failures can be subtle and consequential, instructors must validate the feedback generated before it reaches students. 
However, this validation work can be tedious and repetitive~\cite{tang2025sphere} across large student submissions where instructors repeatedly encounter recurring error patterns and pedagogical situations that call for similar judgments, but with important contextual differences.

Prior work has explored several ways to make human validation of AI-generated content reusable at scale. Some systems ask users to explicitly specify evaluation criteria or policies\cite{kim2024evallm,lam2025policy,gao2025llm,zheng2023judging}, while others infer criteria or model behavior from users’ judgments on generated outputs\cite{shankar2024validates,shaikh2024aligning}. 
These approaches reduce repeated review, but they also expose a tradeoff between specification effort and situated judgment: criteria written upfront ask users to anticipate relevant cases, whereas judgments made on concrete outputs are easier to provide but may not produce an explicit specification that users can inspect and reuse.\cite{shankar2024validates,shaikh2024aligning}

To understand how this tradeoff manifests in programming education, we first conducted an observational study of instructors validating AI-generated feedback without criterion-authoring support. Across 178 validation episodes, instructors frequently (69.66\%) validated feedback in relation to the corresponding student submission and revised it based on that context. Yet those decisions remained local to individual cases, requiring similar judgments to be reconstructed when related situations appeared elsewhere.
This structure led us to recast scalable feedback validation as an \textit{end-user programming} problem, where the goal is to capture and reuse the logic behind instructors’ situated judgments.

Specifically, when instructors validate feedback, they are implicitly expressing conditional logic: when a particular situation occurs in the student submission and generated feedback, the feedback should change in a particular way. Scaling that judgment therefore requires activities similar to programming, including specifying conditions and actions, applying them to new cases, inspecting their behavior, and revising them when they fail. The challenge is to give instructors this programmatic leverage without requiring them to stop their pedagogical work and write a complete program from scratch.

Programming-by-Demonstration (PbD) offers one way to support this kind of end-user programming where users demonstrate desired behavior on concrete examples, and the system infers reusable behavior from those demonstrations\cite{cypher1993watch}. Feedback validation, however, poses a particular challenge.
The demonstrated action itself, such as revising a feedback message, is easy to perform, but the reusable abstraction behind that action is not necessarily apparent.
The system must infer when the same pedagogical judgment should apply, while instructors must be able to inspect and refine where that abstraction generalizes. 
Therefore, we ask: \textbf{\textit{How can we scale instructors’ validation of AI-generated feedback by turning their pedagogical judgments into reusable logic that they can understand and refine across diverse student submissions?}}

In this paper, we introduce \sys{}, a feedback validation system that turns instructors’ validation behaviors into reusable, inspectable programs.
We call this approach \textit{validation by demonstration} (VbD).
\sys{} treats an instructor’s revision of AI-generated feedback as a demonstration of an underlying validation criterion, infers reusable pedagogical logic from that demonstration, and allows instructors to examine how the inferred logic behaves across heterogeneous student submissions. 
We iteratively designed \sys{} through a comparative formative study to better align this representation with instructors’ validation workflows while keeping the resulting criteria easy to understand and refine.

We evaluated \sys{} in a within-subjects study with 16 instructors experienced in programming and teaching. Compared with a natural-language baseline, participants using \sys{} created criteria covering more distinct pedagogical intents ($\mu = 2.56$ vs. $2.00$, $p<.05$); the resulting feedback satisfied a greater proportion of instructors’ own criteria (91\% vs. 57\%, $p < .01$) and more than doubled their accuracy in correcting problematic criteria (77\% vs. 32\%, $p < .0001$). 
System interaction logs further showed that instructors did not simply specify criteria once and apply them at scale; instead, they repeatedly applied, inspected, and refined criteria across new submissions. 
These findings suggest that \sys{} can help instructors not only turn their judgments into explicit, reusable criteria, but also progressively clarify those criteria by seeing where they did and did not apply across other submissions.
This work points to a form of human-AI augmentation in which systems help experts make partially articulated knowledge visible, testable, and reusable.

Our research makes the following contributions:
\begin{itemize}
    \item Design implications, drawing insights from a formative study, for supporting users in validating large-scale LLM-generated programming feedback by demonstration.
    \item A new interaction technique for scalable LLM-generated programming feedback validation, \textit{validation-by-demonstration}, that infers users' validation intentions and represents them in a new DSL for users to flexibly review, adjust, and apply to all other feedback.
    \item An interactive system, \sys{}, that implements our validation-by-demonstration approach with a mixed-initiative interface for effective human-AI co-validation in validating programming feedback at scale.
\end{itemize}

\section{Related Work}

\subsection{LLM-assisted Evaluation}
Though presenting generated documents with similarity-based visual augmentation can support users' sensemaking process~\cite{SensemakingLLMGero2025}, validating large-scale generated texts requires complex attention shifts that can induce decision fatigue and lead to an infeasible workload. As LLM and LLM-as-a-judge~\cite{zheng2023judging} techniques advance, a growing body of AI-assisted interactive evaluation tools has focused on different aspects of the evaluation process, such as facilitating prompt iteration or refining evaluation criteria~\cite {ashktorab2024aligning, arawjo2024chainforge}. These systems highlight LLM's power in supporting users to handle large-scale data, despite the substantial needs for users to efficiently specify their intentions, identify, and repair human-LLM misalignments. For example, EvalLM~\cite{kim2024evallm} explored methods to help users compose criteria, and EvalGen~\cite{shankar2024validates} used users' ratings of generation results to synthesize validation criteria. To enable users to map concepts from validation criteria to actual data points, Policy Maps~\cite{lam2025policy} defined structured criteria and generated concept components beforehand based on the dataset's characteristics, and Evalet~\cite{kim2025evalet} divided generated content into meaningful chunks and generated functions to produce traceable evaluation. Building on these, our approach extracts validation criteria that align with user intention from their demonstration of revising programming feedback, amends the criteria by maintaining instructors' user model, and visualizes text-span level concept matches across code and feedback.

\begin{table}[t]
  \caption{\textbf{Design space for generative AI pedagogical guidance validation.}We treat \emph{validation} as any instructor act that governs whether generated guidance reaches learners.
  \textbf{Rows} name that act: \textsc{state} a criterion before seeing output, \textsc{rate} outputs presented for review, or \textsc{demonstrate} by revising an output in place.
  \textbf{Columns} name what the act leaves behind for outputs the instructor never sees: \textsc{none} (applies only to the item touched), \textsc{opaque} (reused by the system but not inspectable), \textsc{readable} (natural language the instructor can rewrite), and \textsc{structured} (an executable specification the instructor can inspect and repair).
  \textit{n/a} marks cells that cannot occur, since a stated criterion is itself the persisted, readable artifact.
  $\bigcirc$ marks cells with no published occupant known to us.}
  \label{tab:designspace}
  \small
  \setlength{\tabcolsep}{6pt}
  \renewcommand{\arraystretch}{1.35}
  \begin{tabular}{@{}l !{\color{black!50}\vrule width 0.7pt} *{4}{>{\raggedright}p{0.182\linewidth}}@{}}
    \toprule
    \textbf{Act}
    & \textsc{None} & \textsc{Opaque} & \textsc{Readable} & \textsc{Structured} \tabularnewline
    \midrule
    \rowcolor{black!6}
    \textsc{State}
    & \textcolor{black!45}{\textit{n/a}}
    & \textcolor{black!45}{\textit{n/a}}
    & CS50 Duck~\cite{liu2024teachingcs50} 
    & CodeHelp~\cite{liffiton2023codehelp}\newline
    Co-Pilot~\cite{hedderich2024piece}\newline
    TeachTune~\cite{jin2025teachtune}
    \tabularnewline
    \textsc{Rate}
    & SPHERE~\cite{tang2025sphere}\newline Edison~\cite{miroyan2025edison}
    & CS50 human feedback~\cite{liu2025improvingcs50}\newline DPO with teachers~\cite{woodrow2025dpo}
    & GradeHITL~\cite{chu2025gradehitl}
    & \textcolor{black!45}{$\bigcirc$}
    \tabularnewline
    \rowcolor{black!6}
    \textsc{Demonstrate}
    & Teacher revisions~\cite{borchers2026understanding}
    & CoGrader~\cite{chen2025cograder}
    & Avalon~\cite{armfield2025avalon}\newline \textit{Formative baseline}~(\S\ref{sec:formative})
    & \textbf{\sys{}}
    \tabularnewline
    \bottomrule
  \end{tabular}
\end{table}

\subsection{Pedagogical LLM-Generated Feedback Validation}
\label{subsec:rw-validation}

Pedagogical feedback is among the most influential factors in learning outcomes~\cite{hattie2007power}. LLMs can generate individualized programming feedback~\cite{nguyen2024using, nguyen2024comparing, gabbay2024combining}, with some approaches incorporating teacher supervision~\cite{woodrow2025dpo}. LLM-generated guidance is already reaching students in programming courses~\cite{liu2024teachingcs50, liffiton2023codehelp}. However, its reliability remains uneven~\cite{estevez2024evaluation}, and educators' concerns extend beyond incorrect outputs~\cite{harvey2025don}.
Existing systems offer different ways for instructors to control the guidance students receive. Some constrain generation before outputs are produced: CS50 Duck~\cite{liu2024teachingcs50} uses written instructions, TeachTune~\cite{jin2025teachtune} represents instructions as a node graph, and CodeHelp~\cite{liffiton2023codehelp} and Hedderich et al.'s Co-Pilot~\cite{hedderich2024piece} use structured configuration. Others ask instructors or teaching assistants to rate generated outputs. SPHERE~\cite{tang2025sphere} and Edison~\cite{miroyan2025edison} collect judgments without reusing them. In contrast, CS50's human-feedback loop~\cite{liu2025improvingcs50} and a teacher-in-the-loop DPO pipeline~\cite{woodrow2025dpo} incorporate ratings into the model. GradeHITL~\cite{chu2025gradehitl} instead revises a written rubric based on instructors' answers to its questions. A third group asks instructors to revise outputs directly, whether to study teachers' revisions~\cite{borchers2026understanding}, steer LLM assessment of the remaining reports~\cite{chen2025cograder}, or calibrate an LLM grader~\cite{armfield2025avalon}. Table~\ref{tab:designspace} organizes these systems by the act the instructor performs and what that act leaves behind to guide outputs the instructor does not inspect.

Each region of this space carries a cost. Stating a criterion upfront asks instructors to articulate in advance the cases that only outputs would reveal~\cite{liffiton2023codehelp, liu2024teachingcs50}. Folding judgments into opaque model updates generalizes with almost no added burden~\cite{liu2025improvingcs50, woodrow2025dpo}, but leaves no artifact the instructor can read, contest, or reuse. Revising in place needs no articulation beyond the revision, but its effect stops at the item touched~\cite{borchers2026understanding}. The space therefore trades articulation burden against retained control. Rating leaves that trade-off unresolved because its signal is scalar: knowing that an output is unsatisfactory does not say which condition produced it. A readable criterion then needs a second channel such as GradeHITL's questions~\cite{chu2025gradehitl}, and only outside education has rating reached the structured column~\cite{shankar2024validates}. A revision carries more of the signal, since it shows what should change and where, though the condition under which the change should recur is still left to inference (\S\ref{subsec:rw-pbd}). Demonstration also left the representation question open: in our formative study (\S\ref{sec:formative}), where both conditions allowed criteria to be created by demonstration, instructors produced a significantly higher share of redundant criteria with natural-language criteria than with structured \texttt{when}-\texttt{then} criteria. \sys{} occupies the demonstrate-structured cell: the instructor demonstrates on a single output, and the system externalizes that demonstration as a structured, evidence-grounded criterion the instructor can inspect, repair, and apply across the corpus.

\subsection{AI-assisted Educational Systems}
Recent work has increasingly applied large language models (LLMs) to educational contexts, supporting both student learning and instructor practices~\cite{chen2024learning, jin2024teach}. Here, we focus on the latter: systems that assist instructors with tasks about and beyond providing feedback. For instance, LLMs have been employed in course planning~\cite{fan2024lessonplanner}, assignment and project design~\cite{ravi2025co}. Other systems use LLMs for cross-disciplinary idea generation~\cite{fan2025litlinker} or to simulate virtual peers and teaching assistants that enrich classroom interactions and collaborative learning~\cite{liu2024classmeta,kazemitabaar2024codeaid}. LLMs have also been integrated into learning analytics pipelines and formative feedback systems to deliver personalized support grounded in students' behavior and performance~\cite{zhang2025cpvis}. Collectively, these studies underscore the expanding role of LLM-based tools in instructional design, classroom facilitation, and teaching support. At the same time, emerging findings caution that such systems may risk diminishing instructor engagement and cognitive involvement over time~\cite{lee2025impact,kumar2024human}, further highlighting the importance of keeping instructors in the loop when designing intelligent educational systems.

\subsection{Programming by Demonstration}
\label{subsec:rw-pbd}
Programming by demonstration (PbD) and programming by example (PbE) infer a program from a user's actions or examples on concrete cases~\cite{cypher1993watch,lieberman2001your}. Because a few demonstrations are consistent with many programs, each system needs a way to choose one that generalizes as intended. FlashFill~\cite{gulwani2011automating} and SMARTedit~\cite{lau2003programming} represent the programs consistent with the examples and rank them, Wrangler~\cite{WranglerKandel2011} and SUGILITE~\cite{li2017sugilite} return the ambiguity to the user through previews and confirmations, and Blue-Pencil~\cite{miltner2019fly} recovers repeated edits from a noisy, partial edit stream. In programming education, FixPropagator~\cite{head2017writing} learns a code transformation from a teacher's fix to one submission and propagates the teacher's written feedback to every submission that transformation fixes. \sys{} follows this line of work in inferring a rule from an action on one case, and, as in PbE, one example under-determines the rule. The difference is what must be inferred: unlike a code fix, a feedback revision shows what should change but not when it should recur, and that condition, over the untouched submission and in semantic labels an LLM judges at execution, cannot be checked symbolically, so the instructor inspects where the rule holds (\S\ref{subsec:dslRule}).

\begin{table*}[t]
  \footnotesize
  \setlength{\tabcolsep}{5pt}
  \renewcommand{\arraystretch}{1.5}
  \begin{tabular}{@{}>{\raggedright\arraybackslash}m{0.11\linewidth}
                    >{\raggedright\arraybackslash}m{0.20\linewidth}
                    >{\raggedright\arraybackslash}m{0.30\linewidth}
                    >{\raggedright\arraybackslash}m{0.19\linewidth}
                    >{\raggedright\arraybackslash}m{0.10\linewidth}@{}}
    \toprule
    \textbf{Workflow} & & \textbf{Definition} & \textbf{Example sequence} & \textbf{\% (n)} \\
    \midrule
    \textit{Feedback-only}
      & \includegraphics[height=30pt]{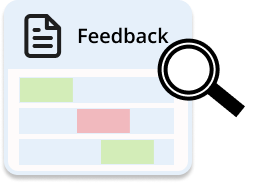}
      & The instructor settles the generated message without opening the submission.
      & read message $\rightarrow$ send
      & 30.34 (54) \\
    \midrule
    \textit{Code-first}
      & \includegraphics[height=30pt]{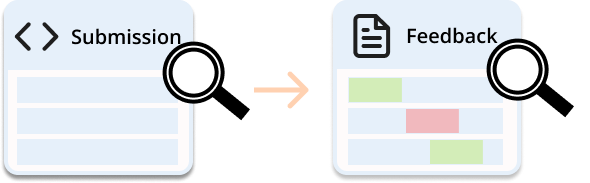}
      & The instructor reads the submission first and forms their own judgment of what the
        student needs, then turns to the message and decides against that judgment without
        going back.
      & read submission $\rightarrow$ read message $\rightarrow$ send
      & 11.80 (21) \\
    \midrule
    \textit{Interleaved}
      & \includegraphics[height=34.75pt]{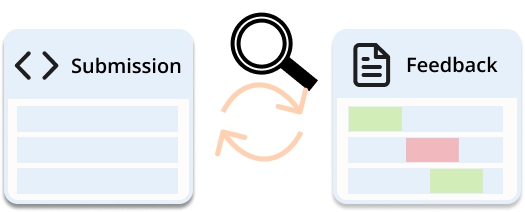}
      & The instructor enters from either the submission or the message, then keeps moving
        between the two, revising the feedback and checking it against the code again as
        they go.
      & read message $\rightarrow$ check submission $\rightarrow$ revise message
        $\rightarrow$ check submission $\rightarrow$ send
      & 57.87 (103) \\
    \bottomrule
  \end{tabular}
  \caption{\textbf{Validation workflows from the preliminary observations.} The last column
  gives each workflow's share of the 178 coded episodes, with the episode count in
  parentheses. Episodes are coded one at a time, so an instructor can appear in more than one
  workflow, and none of the ten stayed with a single workflow.}
  \label{tab:workflow-types}
\end{table*}

\section{Formative Study}
\label{sec:formative}
\subsection{Preliminary Observational Study}
\label{sec:Preliminary}
To explore the design space (Table~\ref {tab:designspace}) and to understand how instructors' natural behaviors and workflows on validating LLM-generated programming feedback when no criteria authoring support is available, we conducted a preliminary study with 10 participants (5 male, 5 female) who have both programming and teaching experience. Participants were asked to review LLM-generated feedback for students working on an introductory Python programming exercise and send messages they choose with or without revision for 12 minutes. Each generated message was displayed next to the student's submission, and participants could edit the message or adjust the generation prompt before sending or discarding it. The interface gave them no way to record a validation criterion or apply it to other students, so the validation logic we observed is their own. Each study session lasted about 20 minutes and was conducted in person. Participants were asked to think aloud, and the sessions were video and screen recorded.

To analyze the results, we coded \textit{validation episodes} from the screen recordings, running from the participant first engaging a generated message to sending it, modifying it, or leaving it (178 episodes across the 10 sessions, $\mu=17.80$ per participant). We use cursor position, text selection, control clicks inside each panel, and their think-aloud as a proxy for where the participant was working. Since the submission and the feedback were on screen together, we cannot establish gaze, and a feedback-only episode should be read as one in which the participant did not interact with the submission, rather than one in which they did not read it.

We observed three ways participants worked for feedback validation (Table~\ref{tab:workflow-types}). In \textit{feedback-only} episodes ($n=54$, $30.34\%$), the participant settled the generated message without interacting with the submission. In \textit{code-first} episodes ($n=21$, $11.80\%$), they worked through the submission and formed their own reading of it before turning to the message. In \textit{interleaved} episodes ($n=103$, $57.87\%$), participants entered from either the submission or the message and kept moving between the two, revising the feedback and checking it against the code again as they went. we define a participant's interleaving rate as $\frac{\# of Interleaved Episodes}{\# of Validation Episodes}$ ($\mu=0.60$, $\sigma=0.15$, ranging from $0.35$ to $0.91$). These are codes on single decisions, and no participant stayed with one workflow for a whole session. Every participant produced at least one interleaved episode, 9 of 10 produced at least one feedback-only episode, and 7 of 10 produced at least one code-first episode.

What participants did in the latter two cases was judging a message against the submission it was attached to, and the same message could be right for one submission and wrong for another. PP1 revised feedback where \textit{the response that the AI generated was kind of not accurate \ldots it didn't really make sense for the code that was there},'' and PP3 rewrote messages that gave \textit{not just [un]necessary information},'' but ``\textit{did not point out the key mistake they made}.'' Participants made judgments of this kind repeatedly as they worked through the batch, and the interface offered no way to record one and reapply it, so every submission had to be decided by hand. This suggested a design opportunity to capture what an instructor decides in a single case and carry it to the submissions they have not reached, as long as what gets captured can refer to the submission and the feedback together.

\begin{figure}[t]
    \centering
    \includegraphics[width=1\linewidth]{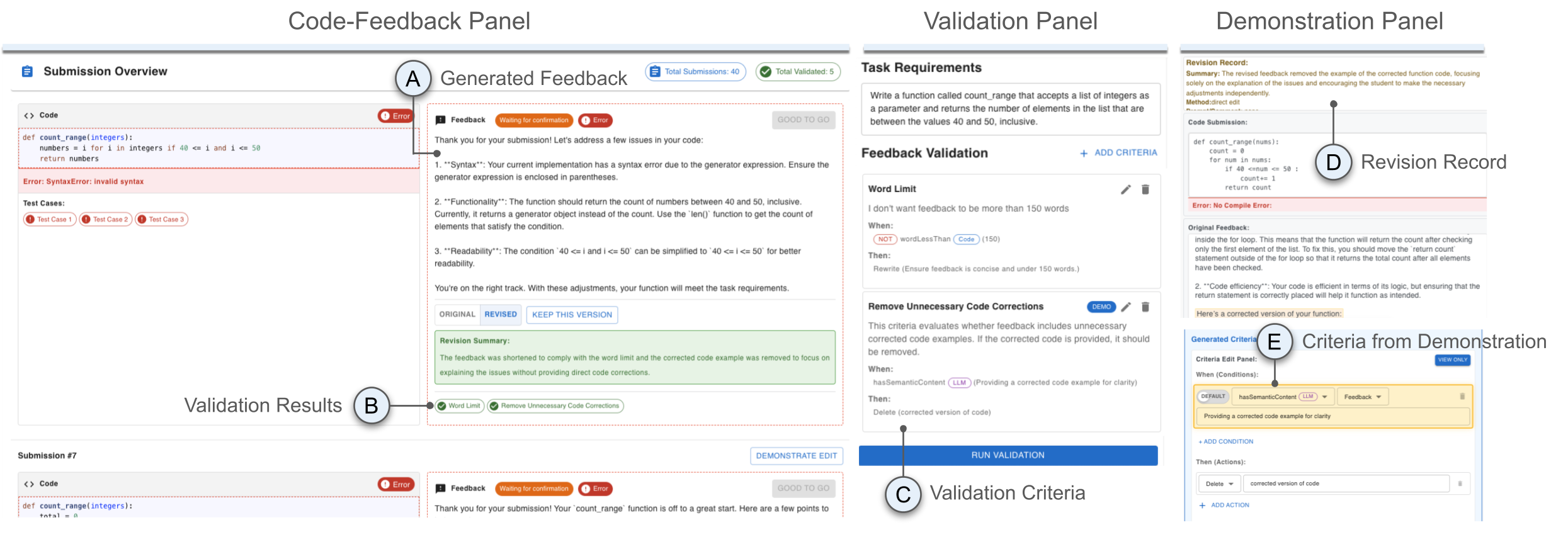}
    \caption{\textbf{Probe system} is composed of three main panels: code-feedback, validation, and demonstration. In the code-feedback panel, the instructor can review the generated feedback (A), the system's validation results (B), and validate and edit feedback. In the validation panel, the instructor can create, manage, and apply validation criteria (C). In the demonstration panel, the instructor can review their revision record (D), and adjust the criteria generated from their demonstration (E). }
    \label{fig:probe-UI}
\end{figure}

\subsection{Design Probe}
\label{sec:prob}

To explore whether such a decision can be captured and carried forward, we developed a system probe based on insights drawn from the preliminary observation and prior work. The probe let instructors revise generated feedback, inferred a structured \texttt{WHEN}-\texttt{THEN} criterion from that revision, and applied the criterion across the remaining submissions for review. The probe serves as a pilot artifact used to surface breakdowns in criterion understanding and verification. 

Two observations from the preliminary study (\S\ref{sec:Preliminary}) shaped the probe's criteria format. In the code-first and interleaved workflows instructors reached their decision against the submission, so a criterion has to be able to refer to the submission as well as the message. What instructors then did about a misaligned message was revise it rather than score it, so a criterion also has to carry the edit and not only the judgment.

Since instructors moved between the two surfaces while deciding, the generated feedback is displayed side-by-side with the submission (Fig.\ref{fig:probe-UI}A) so that they can cross-check without navigating away. The validation criterion is therefore presented in a two-phase structured format (Fig.\ref{fig:probe-UI}C), where the \texttt{WHEN} statement represents the conditions to be matched over the submission and the feedback, and the \texttt{THEN} statement shows the revision to be applied (probe's DSL in Appendix~\S\ref{apx:design_probe}). After a demonstration edit, the probe system generates validation criteria (Fig.\ref{fig:probe-UI}.E) for the instructor to review and refine. After creating validation criteria, the instructor can apply them to all submissions and review the results (Fig.\ref{fig:probe-UI}) to decide which feedback to keep, so that a decision made on one message reaches the submissions they have not opened.

To generate validation criteria from user's demonstration, we employ an off-the-shelf LLM (i.e., GPT-4o), to process the relevant demonstration context (specifically: \texttt{programming problem}, \texttt{code submission}, \texttt{user edits log}, \texttt{feedback before revision}, and \texttt{feedback after revision}) with our criteria format.

\subsection{Methodology}
We recruited 12 participants (7 male, 5 female) who have both programming ($\mu = 3.75$ years, $\sigma=1.86$) and teaching experiences ($\mu=2.50$ years, $\sigma=1.24$) to participate in our within-subjects study where they used the probe and a baseline system to validate generated feedback. Participants completed a survey after each condition and a semi-structured interview at the end of the study. We used data from two authentic coding exercises in a local university's introductory-level Python class. The baseline condition replicated the common practice of \textit{LLM-as-a-judge}, where users create validation criteria in pure natural language. To explore the influence of criteria formats, the baseline condition also allowed users to create criteria from their demonstration. In each condition, participants use the system to validate LLM-generated feedback for incorrect student submissions, aligning it with their own teaching style and strategy. 

\begin{table*}[t]
  \footnotesize
  \setlength{\tabcolsep}{5pt}
  \renewcommand{\arraystretch}{1.15}
  \begin{tabular}{@{}c>{\raggedright\arraybackslash}p{0.14\linewidth}>{\raggedright\arraybackslash}p{0.28\linewidth}>{\raggedright\arraybackslash}p{0.36\linewidth}r@{}}
    \toprule
    \textbf{Index} & \textbf{Code} & \textbf{Definition} & \textbf{Example criteria} & \textbf{\% (n)} \\
    \midrule
    \textbf{1} & \texttt{conciseness} & The criterion cuts material that does not teach: greetings, filler, over-long explanation, or encouragement the student does not need. & \emph{``Feedback must exclude any final redundant encouraging message.''}\newline \emph{``Ensure that the feedback is concise and actionable.''} & 33.73 (28) \\
    \addlinespace[3pt]
    \textbf{2} & \texttt{no\_direct\_solutions} & The criterion keeps complete code solutions and direct answers out of the feedback, so the student has to work the problem out. & \emph{``Feedback must not include any source code snippets.''}\newline \emph{``Confirm that the feedback does not provide direct answers.''} & 24.10 (20) \\
    \addlinespace[3pt]
    \textbf{3} & \texttt{content\_requirement} & The criterion asks the feedback to cover something specific, such as the range function, list comprehensions, or testing with different inputs. & \emph{``Inclusivity of Range: explain how range works.''}\newline \emph{``Feedback includes an explanation of the range function if it is missing.''} & 9.64 (8) \\
    \addlinespace[3pt]
    \textbf{4} & \texttt{code\_style} & The criterion asks the feedback to comment on how the code is written, most often its naming. & \emph{``The feedback must emphasize the use of descriptive variable names.''}\newline \emph{``The code lacks clarity in parameter naming. If the parameter naming is unclear, it suggests replacing the function\ldots''} & 6.02 (5) \\
    \addlinespace[3pt]
    \textbf{5} & \texttt{specify\_program} & The criterion states what the student's program itself has to do or look like, rather than what the feedback should say about it. & \emph{``Ensure the input to the make\_list function is a list.''}\newline \emph{``Check whether the function is returning the correct list or not.''} & 6.02 (5) \\
    \addlinespace[3pt]
    \textbf{6} & \texttt{conceptual\_guidance} & The criterion steers the feedback toward the idea behind the mistake instead of the particular fix. & \emph{``Feedback should encourage problem-solving or exploration without providing exact code solutions.''}\newline \emph{``Overly specific variable name suggestions to maintain generality in feedback.''} & 4.82 (4) \\
    \addlinespace[3pt]
    \textbf{7} & \texttt{feedback\_structure} & The criterion sets how the feedback is arranged. Its sections, its ordering, whether it runs as paragraphs or as bullets. & \emph{``Make the feedback one paragraph, avoiding numbered or bullet points.''}\newline \emph{``Feedback must be organized into two distinct categories: `Necessary Changes' and `Bonus Changes'.''} & 4.82 (4) \\
    \addlinespace[3pt]
    \textbf{8} & \texttt{scope\_control} & The criterion takes out advice that falls outside what this submission needs. & \emph{``The feedback contains code style advice and removes it if present, ensuring that feedback does not include\ldots''}\newline \emph{``The feedback contains suggestions for non-essential improvements to the code and, if so, removes those suggestions to\ldots''} & 4.82 (4) \\
    \addlinespace[3pt]
    \textbf{9} & \texttt{error\_explanation} & The criterion asks the feedback to name or explain the error, including the message or the type. & \emph{``The feedback must specifically address syntax errors.''}\newline \emph{``The feedback must include an explanation of the error type encountered.''} & 3.61 (3) \\
    \addlinespace[3pt]
    \textbf{10} & \texttt{positive\_tone} & The criterion asks the feedback to stay warm and supportive. & \emph{``Encourage students.''}\newline \emph{``Maintain a positive and encouraging tone throughout the feedback.''} & 2.41 (2) \\
    \bottomrule
  \end{tabular}
  \caption{\textbf{Pedagogical intent codebook.} Ten codes derived from the 83 validation criteria participants created in the formative study, ordered by frequency. Each row gives the code, its definition, and two example criteria from the corpus.}
  \label{tab:intent-codebook}
\end{table*}

\subsection{Results and Design Considerations}

To analyze the results, we collected user-validated feedback, coded the comments from the semi-structured interviews, ran the Wilcoxon signed-rank test for Likert-scale survey responses, counted the number of validated feedback, created criteria, and ran the t-test. We also coded the criteria created by participants in both conditions. To capture what participants were trying to achieve rather than only measuring the numbers of criteria, two researchers first derive the pedagogical intent codebook independently from the study's criteria corpus ($n=83$), and then meet and merge their codebooks to a shared codebook for user study analysis. After that, both coders independently code 10 criteria randomly sampled from the study corpus, and have another meeting for codebook calibration. Then, two coders independently code 30 random samples (36.14\% of the study corpus) using the final codebook~(Table~\ref{tab:intent-codebook}), achieving Cohen's $\kappa=0.86$, and resolved disagreements through discussion and coded the rest of the corpus independently. Coders were blind to the system conditions , and the probe's \texttt{WHEN-THEN} criteria were translated to natural language format to avoid decision bias.

Both the probe and the baseline scale the validation through the validation criteria. A criterion from the probe matched a mean of 23.6 code-feedback pairs, while a criterion from the baseline matched 18.1.
We did not find a significant difference in the number of feedback user validated (Probe: $\mu=25.50, \sigma=13.00$; Baseline: $\mu=21.92, \sigma=10.9$) and the number of criteria user created (Probe: $\mu=3.50, \sigma=1.44$; Baseline: $\mu=3.42, \sigma=2.15$) across two conditions.
There was not a statistically significant difference in the Likert-scale responses between probe and baseline.
To compare the diversity of criteria, we define that two criteria are in the same aspect if they are semantically identical with only differences in wording, and we further define the efficiency of criteria as $\frac{\# of Unique Aspects}{\# of Criteria}$. We identified a significantly higher efficiency using Probe (Probe: $\mu=0.91, \sigma=0.13$; Baseline: $\mu=0.72, \sigma=0.21, p<0.05$). This happens because in Baseline many participants (FP1, FP2, FP4, FP9, FP10, FP11) created multiple validation criteria with the same intention due to the poor performance of the prior criteria. We found that while in aggregate more than half of the validation criteria were created through demonstration (probe: $64.00\%$, Baseline: $61.00\%$), but the ability to create criteria from demonstration did not always improve participants' experience. For example, FP9 created 9 criteria in total with only efficiency of $0.33$, since they kept creating criteria to enforce a specific level of brevity, but eventually failed due to the lack of prompting skill under the baseline condition. This implies the need to support instructors in managing their validation criteria with respect to redundancy and conflicts (\textbf{DC1}).

Per the pedagogical intent coding results~\ref{tab:intent-codebook}, the dominant top 2 intents are about \texttt{conciseness}~(probe: $n=16$; Baseline: $n=12$) and \texttt{no\_direct\_solution}~(probe: $n=10$; Baseline: $n=10$). Participants' criteria from the Baseline system cover all 10 types of intents while the criteria from the probe cover 8 out of 10 intents, missing \texttt{error\_explanation} and \texttt{positive\_tone}. Meanwhile, we also notice that the current predicates in the \texttt{WHEN-THEN} structure would not support participants' revision intents about substitution. These findings reveal the need for a criterion structure that balances the expressiveness and control (\textbf{DC2}).

From the post-study questionnaire and semi-structured interview, we found that the probe's structured criteria format brought participants high level of perceived feeling of control. 7 out of 12 participants rated the highest score of control using \sys{}. FP4 stated that ``\textit{I had a lot of control (using probe), like, I was able to tell the AI exactly what I wanted to have, and also be able to edit it easily.}'' Likewise, FP5 described the control they perceived using pure natural language criteria as being ``\textit{not as effective as the other.}'' FP8 also expressed their concern when using Baseline that they were ``\textit{not sure how detailed I have to go to get ideal validation.}'' 11 out of 12 participants also stated that they prefer to use the probe system for long-term feedback validation. FP9 also appreciated the flexibility the probe has in managing criteria that ``\textit{I can easily adjust criteria across use cases because of the structure of condition and action, but I don't know what to do to make the correct adjustment in the other condition.}'' 

However, the probe demanded a high level of \textit{cognitive load} as participants needed extra effort to understand the validation criteria. For example, some participants (FP1, FP3, FP10) find it difficult to map the generated criteria back to the demonstration. FP1 stated that they ``\textit{overthink and feel helpless}'' when reviewing structured criteria generated from their demonstration in the probe system. In addition, the probe's structured criteria can fail to maintain users' feelings of control when there is no clear signal and evidence corresponding to the criteria execution results. Participants could only guess the criteria outcome based on the result icons, in a way ``\textit{if it's red, then maybe I'll check again, or if it's green, I can have less attention to it (FP6).}'' This revealed a need for the next system design iteration to connect the original demonstration and criterion execution outcomes to the inferred criterion in an inspectable, evidence-grounded way, so that instructors can both understand how a criterion was derived and verify how it operates in practice (\textbf{DC3}).

\section{Design Goals}
Our formative study suggested that demonstration helped instructors express validation intent in context without specifying abstract criteria upfront. However, once criteria were inferred, instructors needed to understand how they were derived, how they would generalize, and how they would affect feedback in execution, since they remained responsible for judging pedagogical appropriateness. Based on these insights and prior work, we derived the following design goals:

\begin{itemize}
    \item \textbf{\textit{DG1: Support instructor understanding of how criteria reflect their demonstrations.}} Natural-language criteria are flexible, but they can be underspecified for reliable execution and difficult to revise systematically. Our formative study suggested that structured criteria can better support expressing and managing validation logic, but only when instructors can understand where each generated component comes from. Therefore, the system should link criterion conditions and actions back to demonstration traces, such as edited spans, referenced code or feedback evidence, and intermediate prompts.

    \item \textbf{\textit{DG2: Support instructors' criterion evolution and maintenance.}} Validation criteria drifts during validation, and accumulate criteria that may overlap or conflict~\cite{shankar2024validates,vaithilingam2025semantic}. Our formative study also suggested that instructors valued being able to adjust structured criteria more directly than natural-language criteria. Therefore, the system should support efficient criterion editing and management, allowing instructors to revise criteria and identify potential redundancy or contradiction with minimal overhead.

    \item \textbf{\textit{DG3: Support instructor understanding of criterion generalizability}} A major challenge we observed was uncertainty about what a criterion would do when applied beyond the demonstrated case. Since AI-assisted validation should keep instructors in the decision-making role~\cite{harvey2025don}, the system should make criterion execution inspectable by showing where a criterion matches and what edits it would apply. This allows instructors to verify generalization, diagnose failures, and refine criteria continuously.
\end{itemize}

\begin{figure}[t]
    \centering
    \includegraphics[width=1\linewidth]{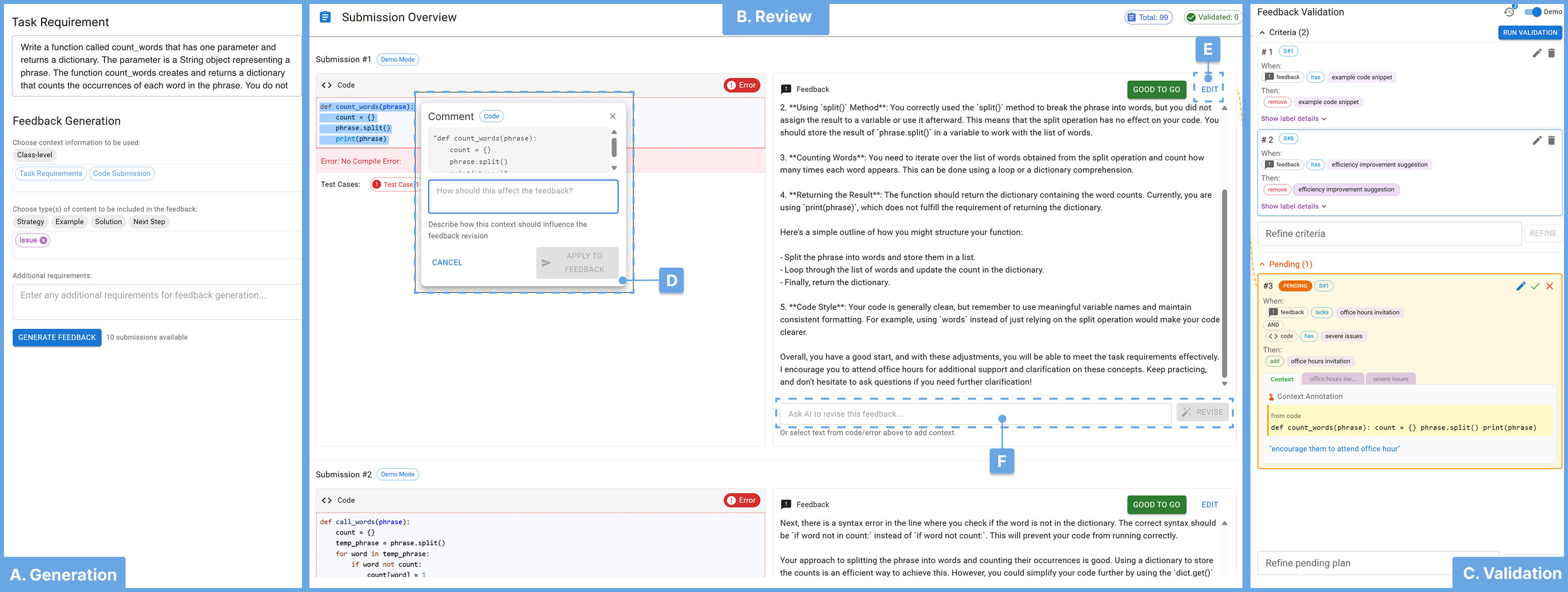}
    \caption{\textbf{\sys{} UI.} \sys{}'s user interface contains three main components: (A) a panel for initial LLM-powered feedback generation, (B) a panel for reviewing code-feedback pairs, and (C) a panel for validation criteria review and application. The instructor can revise the feedback in three ways: (D) comment on code/feedback, (F) AI chat, and (E) direct edit.}
    \label{fig:system_overview}
\end{figure}

\section{\sys{}}
Based on these design goals, we present \sys{} (Fig.\ref{fig:system_overview}), an interactive system that helps instructors validate and pedagogically align LLM-generated programming feedback at scale. 
At a high level, \sys{} first generates a draft set of feedback for student code submissions based on instructor-provided instructions. 
Instructors then review this feedback by editing responses and adjusting prompt to align it with their pedagogical intent.
\sys captures these revision actions as they unfold and distills them into reusable structured criteria, which then be automatically applied to the remaining feedback, reducing the burden of repeated manual validation.
In this section, we will start with a walkthrough of \sys{}'s core features and interfaces, and follow the technical pipeline about how \sys{} suggests criteria based on instructors' demonstrations.

\subsection{Core Features}
\sys{}'s user interface contains three main components: (1) a panel for initial LLM-powered feedback generation (Fig.\ref{fig:system_overview}A), (2) a panel for reviewing code-feedback pairs (Fig.\ref{fig:system_overview}B), and (3) a panel for validation criteria review and application (Fig.\ref{fig:system_overview}C).

Of these, only the feedback generation panel (Fig.\ref{fig:system_overview}A) is carried over from the probe unchanged. The review panel (Fig.\ref{fig:system_overview}B) keeps the probe's side-by-side layout and its three revision modes (Fig.\ref{fig:system_overview}D--F), but differs in two ways. The probe's result icons are replaced by the matching status of each condition, with the evidence spans highlighted in place (Fig.\ref{fig:execution-review}; \textbf{DG3}), and every criterion in the criteria panel (Fig.\ref{fig:system_overview}C) is linked back to the code-feedback pair it was inferred from (\textbf{DC3}, \textbf{DG1}). The probe's separate demonstration panel is removed. Its revision record now sits with the pending criterion, and the instructor adds examples by selecting text spans directly in the code or feedback (Fig.\ref{fig:criteria-UI}E, F) and reviews the matched spans by cluster (Fig.\ref{fig:criteria-UI}G). The criteria panel also checks a pending criterion for conflict, refinement, or duplication before it is accepted (Fig.\ref{fig:criteria-UI}C; \textbf{DC1}, \textbf{DG2}). Underneath, the inference pipeline is new (Algorithm~\ref{alg:criterionInfer}): edits are segmented into separate revision intents, and candidate criteria are executed against the instructor's own revision and ranked by how well they reproduce it. The DSL was revised as well, with a plainer predicate set, one specification form for semantic and exact text, and a \texttt{replace} action the probe lacked (\S\ref{subsec:dslRule}; \textbf{DC2}).

\subsubsection{Feedback Generation}
To initiate the validation workflow, \sys{} generates a draft set of feedback for all incorrect student submissions using a structured feedback model grounded in prior educational research~\cite{hattie2007power}. 
It defines the core components of effective pedagogical feedback.
Instructors can further customize the generation by specifying additional context, such as course policies, tone, or assignment-specific requirements, before producing the initial draft~\cite{tang2025sphere}. This draft batch serves as the raw material instructors refine through demonstration, grounding all subsequent criteria inference in authentic, course-specific feedback.

\begin{examplecallout}
    In an in-class coding exercise, many of Anna's students either failed the test cases or could not get their code running. Although Anna wanted to provide each student with high-quality feedback, reviewing everything manually would take all day, while directly using LLM-generated feedback risked hallucinations and misleading guidance. To accelerate this process, she uses \sys{} to upload the submissions and generate an initial customized feedback draft for the class.
\end{examplecallout}

\begin{figure}[t]
  \centering
  \resizebox{0.7\linewidth}{!}{$
  \begin{aligned}
  \mathit{Rule} &::= \mathbf{when:}\;\Phi\;\;\mathbf{then:}\;\Delta \\[2pt]
  \Phi &::= \phi \mid \Phi \land \Phi \mid \Phi \lor \Phi \\[2pt]
  \phi &::= p(t_1,\ldots,t_n) \\[2pt]
  \Delta &::= \delta \\[2pt]
  \delta &::= a(u_1,\ldots,u_m) \\[4pt]
  p &::= \texttt{has}\ \mid\ \texttt{lacks}\ \mid\ \texttt{style}\ \mid\ \texttt{length} \\[2pt]
  a &::= \texttt{add}\ \mid\ \texttt{remove}\ \mid\ \texttt{replace}\ \mid\ \texttt{revise} \\[2pt]
  t &::= s \mid \texttt{code} \mid \texttt{feedback} \mid n \mid o
  \qquad
  u ::= s \\[2pt]
  s &::= x \mid \texttt{exact}(x)
  \qquad
  x ::= \textit{quoted\_text}
  \qquad
  n ::= \textit{integer}
  \qquad
  o ::= {<} \mid {\le} \mid {>} \mid {\ge} \mid {=}
  \end{aligned}
  $}
  \caption{\textbf{Validation Criteria DSL.}
  A rule consists of a condition \(\Phi\) and an edit intention \(\Delta\).
  Conditions are composed from predicates over \texttt{code} and \texttt{feedback}.
  Specifications \(s\) can be either open-ended semantic labels \(x\) or exact text through \(\texttt{exact}(x)\).}
  \label{fig:criteria-dsl}
\end{figure}

\subsubsection{Multi-modal Revision} Aligning with common commercial grading tools' (e.g., Gradescope) design of the focused grading panel, the main panel (Fig.\ref{fig:system_overview}B) of \sys{} is placed on the left for instructor to review the code-feedback pairs. Drawn on prior work for AI-assisted text editing~\cite{ExecutableandVerifiableText-Editing}, \sys{} allows the instructor to revise the feedback in three ways (Fig.\ref{fig:system_overview}): (1) direct edit, (2) directly prompting using AI chat, (3) commenting on texts (i.e., code, error message, or feedback ) to prompt AI. 

\begin{examplecallout}
    Through a quick skim, Anna finds the feedback fits her general requirements, while there could be issues hiding in the details. Anna starts to take a closer look and revise the part misaligned with her teaching strategy.
\end{examplecallout}
\subsubsection{Validation Criteria DSL}
\label{subsec:dslRule}

\begin{figure}[t]
    \centering
    \includegraphics[width=\linewidth]{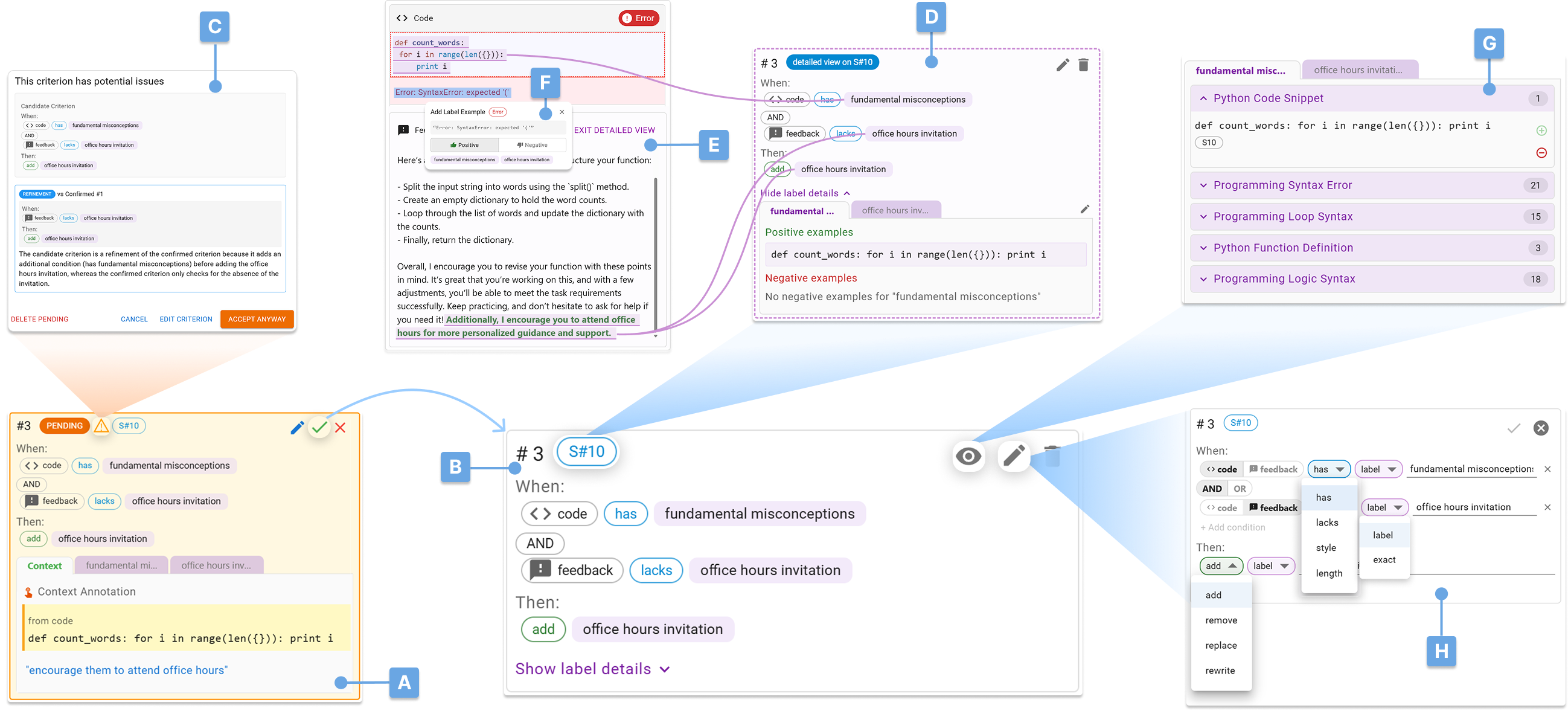}
    \caption{\textbf{\sys{}'s validation criterion} (B). If the instructor attempts to accept a pending criterion (A) with potential issues, a dialog (C) would pop up. The instructor can switch to the detailed view (D) to relate the criterion to the original demonstration, where the semantic labels are linked to the corresponding matched text spans (E). In the detailed view, the instructor can also directly select a text span to add as an example (F). By clicking the view icon, the instructor can overview the matched text spans of semantic labels by clusters and add examples (G). They can also refine the criterion in edit view (H). }
    \label{fig:criteria-UI}
\end{figure}

\sys{} represents validation criteria in a lightweight DSL (Fig.\ref{fig:criteria-dsl}) of the form \(\mathbf{when:}\,\Phi\;\mathbf{then:}\,\Delta\), where \(\Phi\) is a condition over the student's submission and the generated feedback, and \(\Delta\) is an edit intention. Both vocabularies are closed, four predicates and four actions, so an instructor can read the whole language, adjust a criterion directly, and compare two criteria for redundancy or conflict (\textbf{DC1}, \textbf{DC2}, \textbf{DG2}).

The shape of our DSL is designed based on our preliminary observational study (\S\ref{sec:Preliminary}). Participants decided each submission on its own and had no way to record a decision or say which of the remaining submissions it covered, so a rule separates the scope of a judgment (\(\Phi\)) from the change it calls for (\(\Delta\)). What a condition can refer to follows from the three validation workflows (Table~\ref{tab:workflow-types}). Feedback-only episodes ($30.34\%$) produced judgments that hold over the message text alone, such as cutting a closing line of encouragement. Code-first and interleaved episodes ($69.66\%$ together) produced judgments that hold only for a particular pairing of submission and message. Terms therefore range over \texttt{code} as well as \texttt{feedback}, and \(\land\) and \(\lor\) let one rule constrain both at once, as interleaved episodes did when participants moved between the two inside a single decision. Since each predicate is evaluated on its own, execution can later report which part of a condition matched (\textbf{DG3}). \(\Delta\) holds one edit intention, matching the unit we observed: each episode ended in one disposition of one message.

The predicate vocabulary covers the pedagogical intents in participants' criteria (Table~\ref{tab:intent-codebook}). \texttt{has} and \texttt{lacks} state presence and absence, the form most intents take, including \texttt{no\_direct\_solutions}. \texttt{length} takes an integer bound and a comparator for \texttt{conciseness}, the most common intent.
\texttt{style} covers how a message reads rather than what it contains. The actions are \texttt{add}, \texttt{remove}, \texttt{revise}, and \texttt{replace}. The last was added after formative-study participants wrote criteria asking for a substitution that the probe could not express (\textbf{DC2}). Arguments in both clauses share one specification form, \(s ::= x \mid \texttt{exact}(x)\): a bare \(x\) is an open-ended semantic label (e.g., \texttt{"full solution"}) evaluated by the LLM, and \texttt{exact}(x) is literal wording matched programmatically, so an instructor can state which of the two they mean. For example,
\[
\begin{aligned}
\mathbf{when:}\;&\texttt{has(code,"severe issues")} \\
&\quad \land\ \texttt{lacks(feedback,exact("office hour"))} \\
\mathbf{then:}\;&\texttt{add("office hours invitation")}
\end{aligned}
\]
specifies that if the student's code has severe issues and the feedback does not explicitly mention \texttt{"office hour"}, the feedback should be revised to encourage the student to attend office hours.

In the interface, each criterion appears as a structured \texttt{when}-\texttt{then} rule rather than raw code (Fig.\ref{fig:criteria-UI}B), with the condition and action in separate sections, semantic labels visually distinguished from exact text, and positive and negative examples attached to each label. Instructors can read what the system inferred from their demonstration and refine it in place (Fig.\ref{fig:criteria-UI}H), keeping a criterion connected to the evidence behind it (\textbf{DC3}, \textbf{DG1}).

\begin{examplecallout}
     After making several revisions, Anna notices that the tool suggested some validation criteria based on her edits. \sys{} presents the criteria in a structured, programming-block-like way that Anna can simply read through to understand the meaning. 
\end{examplecallout}

\subsubsection{Inferring Reusable Criteria from Instructor Edits}
Instructors in our formative study struggled to trust inferred criteria because they couldn't see where each component came from (\textbf{DG1}).
\sys{} addresses this by inferring validation criteria from demonstration edits and visually grounds them in their source demonstration.
This is inspired by the programming by demonstration approach~\cite{cypher1993watch}: users more reliably convey intent by acting on concrete examples than by authoring abstract rules upfront. After each feedback revision, \sys{} infers one or more validation criteria based on interaction and context information, pending instructor review. A pending criterion is visually linked to the source code-feedback pair, displayed alongside the associated demonstration record and semantic label examples (Fig.\ref{fig:system_overview}C). To allow the instructor to seamlessly relate the inferred criterion to their demonstration when evaluating a criterion, \sys{} also visually links the semantic label back to the exact text span, highlighting the background in the detailed view (Fig.\ref{fig:criteria-UI}D) (\textbf{DG1}, \textbf{DG3}). The instructor can also directly select code and feedback phrases to add them as positive/negative examples (Fig.\ref{fig:criteria-UI}F).

\begin{figure}[t]
    \centering
    \includegraphics[width=1\linewidth]{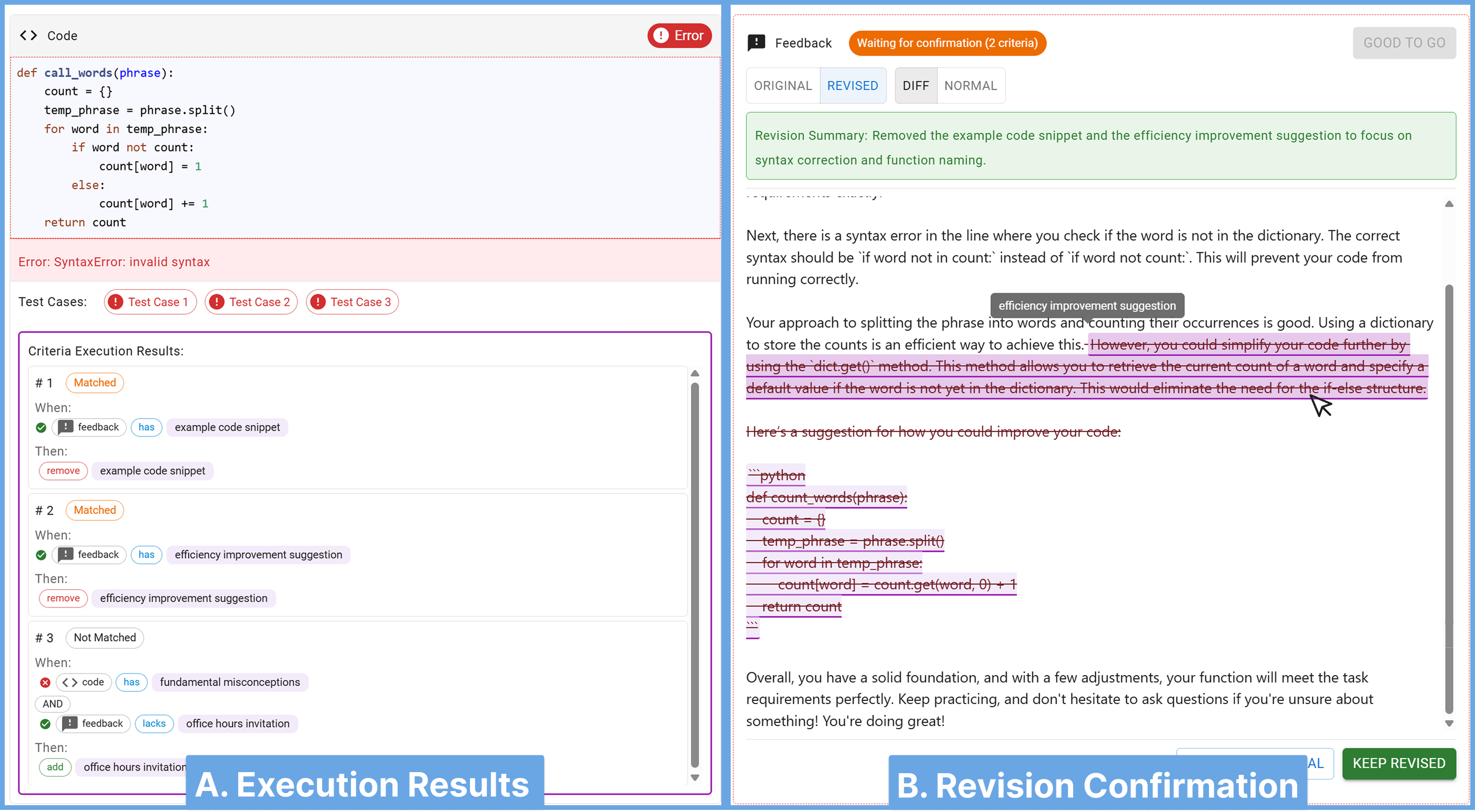}
    \caption{\textbf{\sys{}'s criteria execution UI.} The instructor can inspect the criteria execution results (A) for each code-feedback pair. The matched text spans are highlighted (B) and instructor can hover on a span to view the corresponding semantic label before confirming the revision.}
    \label{fig:execution-review}
\end{figure}

\begin{examplecallout}
Anna notices that one suggested criterion about encouraging students with very flawed code to attend the office hours might be exactly what she wants. She seeks to take a closer look at the details before accepting the suggestions. By interacting with the examples of the semantic labels (``severe issues'' and ``office hours invitation'') and linking them to the code snippet and edits, Anna now has a thorough understanding of what the criterion means, and accepts it to scale the feedback validation process.
\end{examplecallout}

\subsubsection{Detecting Criterion Drift and Redundancy}
When the number of criteria increases, instructors and the probe system could produce criteria that overlap or even conflict with each other, which is a breakdown found in our formative study.
To support instructors in efficiently recognizing and resolving conflicts and evolving the criteria (\textbf{DG2}), \sys{} detects underlying issues across criteria and notifies the instructor of forthcoming conflicts. Whenever a new criterion is suggested, \sys{} would also check if the criterion is a conflict, refinement, or duplicate of existing confirmed criteria. For exact duplicates, the pending criterion would be removed without notifying the instructor. When the instructor accepts a pending criterion with a potential issue, \sys{} would pop up a dialog explaining which confirmed criterion the current criterion might conflict with and why, allowing the user to double-check (Fig.\ref{fig:criteria-UI}C).

\begin{examplecallout}
After accepting several criteria, Anna encounters a pending one that conflicts with an existing confirmed criterion. In the conflict dialog, \sys{} identifies the new criterion,``replace the code example with pseudocode'', as a refinement of the earlier criterion ``remove code examples.'' Realizing her criteria have evolved, Anna chooses to keep the new one.
\end{examplecallout}

\subsubsection{Criteria Execution for Verifiable Validation}
In our formative study, we observed that instructors had no way to verify whether a criterion would generalize beyond the demonstrated case, leading to hesitation in applying it at scale.
To enable verifiable and interpretable criteria execution experience at scale (\textbf{DG3}), \sys{} displays not only if a criterion is matched to trigger feedback revision, but also the intermediate matching status for each condition statement (Fig.\ref{fig:execution-review}A). Meanwhile, for each matched text span, the instructor can view the corresponding semantic label by hovering over the highlighted text (Fig.\ref{fig:execution-review}B). This design aims to empower instructors by allowing them to quickly verify why feedback has been revised in their feedback validation process (\textbf{DG3}). Moreover, \sys{} also enables the instructor to zoom out to review the matched text spans of semantic labels by clusters, so that they can add positive/negative examples from the clusters to better define a semantic label (Fig.\ref{fig:criteria-UI}G).
\begin{examplecallout}
After running auto-validation, Anna navigates to the code-feedback panel to look into the validation results. By reviewing the auto-validation with highlighted evidence, Anna finds a slight misalignment that the system did not treat inline code examples as ``code example''. She then adds more positive examples to refine the matching. With \sys{}'s assistance in criteria creation and feedback validation, Anna's primary role shifts from exhausting feedback review and revision to creating new criteria to ground \sys{}'s validation, while having control over the final approval of distribution.
\end{examplecallout}

\subsection{Technical Pipeline}
\subsubsection{Validation Criteria from Demonstration}
Algorithm~\ref{alg:criterionInfer} summarizes our criteria inference pipeline. Based on our think-aloud study, we found that a single user revision may contain multiple revision intents, which can often be distinguished by the locations and temporal intervals of edits. Accordingly, the system first aggregates keystroke-level events into edit sessions and decomposes them into atomic revision intents. For each segment, it reconstructs a \textit{scoped revision} to isolate only the changes relevant to that intent, generates multiple candidate rules, normalizes them into a valid DSL structure, and filters out contradictory candidates whose conditions make their actions impossible or redundant. The remaining candidates are then described and passed to an interaction-aware maximal marginal relevance (MMR) style selection step, where candidate relevance is computed as $\mathrm{rel}(c)=\alpha\,\mathrm{Rel}_{\text{heur}}(c,s)+(1-\alpha)\,\mathrm{Rel}_{\text{LLM}}(c,s)$. Here, $c$ denotes a candidate rule, $s$ denotes the current revision segment, $\mathrm{Rel}_{\text{heur}}$ is a symbolic alignment score based on entity, predicate, and action match, $\mathrm{Rel}_{\text{LLM}}$ is an LLM-estimated relevance score. Candidates are then greedily selected by maximizing $\lambda\,\mathrm{rel}(c)-(1-\lambda)\max_{c' \in S}\mathrm{sim}(c,c')$, where $S$ is the set of already selected candidates, $\mathrm{sim}$ is the similarity between candidate description embeddings.

The selected candidates are then executed on the original feedback, which serves as a verifiable reward. During execution, exact-text and length predicates are evaluated programmatically, while semantic and stylistic predicates are evaluated with the LLM. After condition evaluation, we prune non-triggered branches in disjunctive condition trees so that action generation is grounded only in the conditions that actually fired. Finally, we rank triggered candidates by how well their outputs align with the user's revision using $\mathrm{score}(c)=w_d\,\mathrm{DiffAlign}(c)+(1-w_d)\,\mathrm{SimAlign}(c)$, where $\mathrm{DiffAlign}$ measures word-level edit overlap, $\mathrm{SimAlign}$ measures embedding similarity to the scoped user revision. We additionally apply a prior bias toward candidates that matches the pedagogical persona from the confirmed criteria. The highest-scoring candidate is returned as the inferred criterion for that segment, together with its description and DSL representation.

\subsubsection{Criteria Execution}
In the automatic validation process, as described in sec~\ref{subsec:dslRule}, for each feedback message, \sys{} will first evaluate it with the conditions in the criteria and execute the user-defined actions on the feedback if the evaluation results of corresponding conditions are matched (i.e., \texttt{TRUE}) with respect to the code-feedback pair. During execution, exact-text and length predicates are evaluated programmatically, while semantic and stylistic predicates are evaluated with the LLM. However, due to the nondeterministic nature of LLM, the actions may not be conducted as described in the actions. To amend this issue, after the feedback has been revised by LLM accordingly, \sys{} will again evaluate the revised feedback message with the conditions, and run actions of those violated rules again, repeating the above until either all conditions are evaluated to be \texttt{FALSE}, or the number of iterations has exceeded the threshold.
\subsection{Implementation}
\sys{} was implemented as a web-based tool using React.js (frontend) and FastAPI(backend). We used OpenAI's GPT-4o~\cite{OpenAIGPT4o} for core LLM functionalities, text-embedding-3-small~\cite{OpenAIEmbedding} for text-embeddings, and AgglomerativeClustering for embeddings clustering.  We also adopted well-known prompting engineering techniques, such as few-shot prompts~\cite{NEURIPS2020_1457c0d6}, AI chains~\cite{10.1145/3491102.3517582}, and persona settings~\cite{10.1145/3586183.3606763}. Hallucination risks in automatic validation are reduced by the design of structural, DSL-based validation criteria, which mixes both \textit{code-based} and \textit{LLM-based} predicates and follows a strict structure and logic.
\begin{algorithm}[t]
\caption{Criterion Inference from an Atomic Revision Intent}
\label{alg:criterionInfer}
\begin{algorithmic}[1]
\Require atomic intent $s$, revision record $D$, confirmed criteria $P$, scoped revision $r_s$

\State generate candidate rules for $s$
\State select a relevant, diverse subset of candidates
\ForEach{selected candidate $c$}
    \State evaluate $c$ on the original feedback
    \If{$c$ triggers}
        \State prune inactive condition branches
        \State execute $c$ to obtain revised feedback $f_c$
        \State score $c$ by edit overlap and semantic similarity to $r_s$
        \State apply a prior bias if $c$ aligns with $P$
    \EndIf
\EndFor
\State return the highest-scoring candidate
\end{algorithmic}
\end{algorithm}

\section{Technical Evaluation}
We compare our validation-by-demonstration pipeline with an existing approach that validates text documents with pure natural-language criteria in a programming feedback validation context:
\begin{itemize}
    \item \textbf{\texttt{Ours:}} We used the approach from our \textit{validation-by-demonstration} pipeline for structured criteria inference from demonstration, and criteria execution.
    \item \textbf{\texttt{NL:}} We adopted the approach and prompts from prior work for LLM-assisted evaluation~\cite{kim2024evallm}, and refined it to fit our context.
\end{itemize}
For both approaches, we used GPT-4o with a temperature of 0.

\subsection{Dataset}
\label{sec:TechEvalDataset}
Two authors who are experienced in both Python programming and programming education annotated two authentic coding exercise datasets to create the technical evaluation dataset. The annotation tool's UI can be found in the Appendix~\ref{apd: techeval}. We used the feedback generation approach in the probe system (Sec.\ref{sec:prob}) to create feedback for each error submission and stored it for static reuse. Two annotators first independently review the LLM-generated feedback and revise the feedback with different methods (i.e., direct edit, prompt, and comment). Annotators were instructed to produce revisions that reflect diverse teaching strategies to lower the risk of biased data collection. To get authentic latent pedagogical intent of their revision, annotators were asked to clearly describe their intent in natural language after they finished a revision. This natural language description was then logged as the ground truth intent for the corresponding revision. Each intent description was cross-checked to ensure clarity.  Each annotator records 10 revisions for each coding exercise dataset, producing 40 revision-intent pairs. After that, we sampled 6 code-feedback pairs for each revision-intent pair as a testing subset, producing 240 test cases in total. Each test case was then annotated to produce the ground truth for criteria matching (i.e., whether the corresponding intent could trigger a revision in the sampled code-feedback pair), by the original creator of this pedagogical intent for authenticity. To mitigate the influence of unbalanced ground truth for criteria matching, annotators manually made sure that there are both positive and negative (i.e., \texttt{TRUE} and \texttt{FALSE}) test cases in each testing subsets.

\subsection{Criteria Suggestion from Demonstration}
In \sys{}, before executing criteria to scale the validation process, the system has to suggest criteria based on users' demonstration. As annotators' intent description already serves as \texttt{NL} approach's criteria, we only ran \texttt{Ours} to generate criteria by inputting context information and user edits from the dataset. For generated criterion, two annotators individually rated the criterion in regards of its clarity and correctness in a Likert-scale format (i.e., 1-5; higher is better). We used Cohen's Kappa to assess the inter-rater reliability. The result show that the criterion generated via \texttt{Ours} approach received average clarity score of 4.34 ($\kappa=0.82$) and correctness score of 4.17 ($\kappa=0.63$).

\subsection{Criteria Execution}
We separated the criteria execution evaluation into two parts: \textbf{criteria matching} and \textbf{criteria action}. Criteria generated in the prior criteria suggestions evaluation were stored and used in this phase of evaluation.

\subsubsection{Criteria Matching}
As stated in the Section~\ref{sec:TechEvalDataset}, we defined that a criterion and a code-feedback pair are matched (i.e., return \texttt{TRUE}) if the feedback needs to be modified to meet the requirements carried in the ground truth intent. For each code-feedback test case, we ran \texttt{Ours} approach with prior generated criterion, and \texttt{NL} approach with the ground truth intent. Over 240 test cases, we observed that \texttt{Ours} ($Accuracy=0.68, Precision=0.72, Recall=0.69$) achieved comparable performance against \texttt{NL} ($Accuracy=0.65, Precision=0.61, Recall=0.99$). Note that the \texttt{NL} used ground truth intents while \texttt{Ours} inferred criteria from demonstration.

\subsubsection{Criteria Action}
We only evaluated \texttt{Ours} approach's criteria action (i.e., the revision quality when feedback needs to be modified), as only \texttt{Ours} approach supports content modification. We randomly sampled 44 of the 89 test cases correctly identified as matched
by \texttt{Ours} ($49.44\%$) and used the produced feedback
revisions as predicted revisions. Then, two annotators individually rated the predicted revisions' correctness in Likert scale (1-5; higher is better) by referring to the original demonstration and recorded intent. The result show that the revisions produced by \texttt{Ours} approach received an average correctness score of 4.73 ($\kappa=0.63$).

\subsubsection{Robustness Evaluation}
Due to the non-deterministic nature of LLM components used in \texttt{Ours}, though augmented by its neuro-symbolic framework, the pipeline cannot guarantee the execution results of the criteria stay the same for multiple runs on the same code-feedback pairs. To measure the multi-run consistency of \texttt{Ours}, we randomly sampled the 10 criteria from the \textit{criteria matching} evaluation, and ran them on the code-feedback pairs from the two coding exercise datasets, respectively 5 criteria on 58 pairs and 5 criteria on 81 pairs for three times (i.e., 695 data points per run, 2085 in total). We used Fleiss' Kappa to assess the consistency of the criteria matching results across runs, and observed a $\kappa$ of $0.81$.

\section{Lab Study}
We conducted a within-subjects user study to evaluate \sys{}'s effectiveness and efficiency in supporting instructors' validation process of LLM-generated programming feedback in regards of criteria creation, application, and correction. The study specifically investigated the following sub-research questions:
\begin{itemize}
    \item \textbf{SRQ1:} Can \textit{validation-by-demonstration} improve the effectiveness, efficiency, and user control during the validation process?
    \item \textbf{SRQ2:} How useful is \sys{}'s structured criteria design in helping user understand and improve their validation criteria?
    \item \textbf{SRQ3:} How would users scale their feedback validation process with \sys{}?
\end{itemize}

\subsection{Participants}
We recruited 16 participants (8 male, 7 female, 1 non-binary) who have both programming ($\mu=3.94$ years, $\sigma=1.29$) and teaching experiences ($\mu=2.38$ years, $\sigma=0.89$) at a local university.

\subsection{Study Procedure}
Each study session lasted about 90 minutes and was conducted fully in person. After receiving a brief introduction to the study's background, participants signed the consent form. Each study participant interacted with \sys{} under two conditions and two datasets. At the beginning of each condition, the study coordinator played the participant a tutorial video of the system. The participant then used the system to complete an LLM-generated feedback validation task. They were asked to align the LLM-generated feedback with their own teaching style and strategy. After each condition, participants completed a post-condition survey, including a NASA-TLX~\cite{hart1988development} (excluding physical demand) survey and Likert-scale questions regarding their experience using the tool. After completing two conditions, the study concluded with a semi-structured interview. All studies were screen- and audio-recorded, and user interaction data were logged through Firebase~\cite{FirebaseUrl}. The orders of conditions and datasets were counterbalanced and randomized.

\subsubsection{Datasets} We used data from two authentic coding exercises in a local university's introductory-level Python course. The Python programming tasks consist of knowledge concepts about function definition, for loop, conditional statement, and list operations. The programming feedback was pre-generated to ensure participants experienced the same feedback.

\subsubsection{Conditions} We used a within-subjects study design where each participant experienced \sys{} under the following two conditions:
\begin{itemize}
    \item \textbf{\sys{}:} a full version of \sys{} with all its features.
    \item \textbf{Baseline:} a baseline version of \sys{} where the validation criteria are in pure natural language. Participants can create criteria by expressing their intent to an LLM and review and refine the generated criteria plan (see Fig.\ref{fig:baseline_user_study} in Appendix). The Baseline system is designed to align with the common \textit{LLM-as-a-Judge} format and people's daily LLM interaction experience.  The system would highlight the relevant text spans for corresponding criteria, instead of semantic labels.
\end{itemize}

\subsubsection{Tasks} For each condition, participant completed the same task of validating LLM-generated feedback from different datasets. There were three sub-tasks that participants completed in order:
\begin{itemize}
    \item \textbf{Criteria Creation} (10 minutes): Participants were provided with a sample dataset including 5 code-feedback pairs, and they were instructed to create validation criteria that match their teaching strategies based on the given data.
    \item \textbf{LLM-assisted Validation} (10 minutes): Participants were instructed to apply the created criteria to scale their feedback validation on a sample dataset including 35 code-feedback pairs. They can edit the feedback if the criteria's execution results did not meet their expectation. They could also create new criteria and modify existing ones, and re-run the criteria. When the feedback is aligned with their teaching strategies, they should mark it as validated.
    \item \textbf{Criteria Correction} (8 minutes): Participants received two criteria that have predefined issues with the criteria execution: code-feedback pairs that should be matched (i.e., feedback should be revised to meet the criteria's standards) are not correctly matched, or vice versa. Each criterion was presented with 4 code-feedback pairs. Participants were instructed to revise the criteria or enrich them with relevant examples to fix the issue, and to re-run if needed for debugging. For each criterion, they have 4 minutes to work on. 
\end{itemize}
At the beginning of each sub-task, the study coordinator would play a tutorial video on the system features relevant to that sub-task.

\begin{figure}[t]
    \centering
    \includegraphics[width=\linewidth]{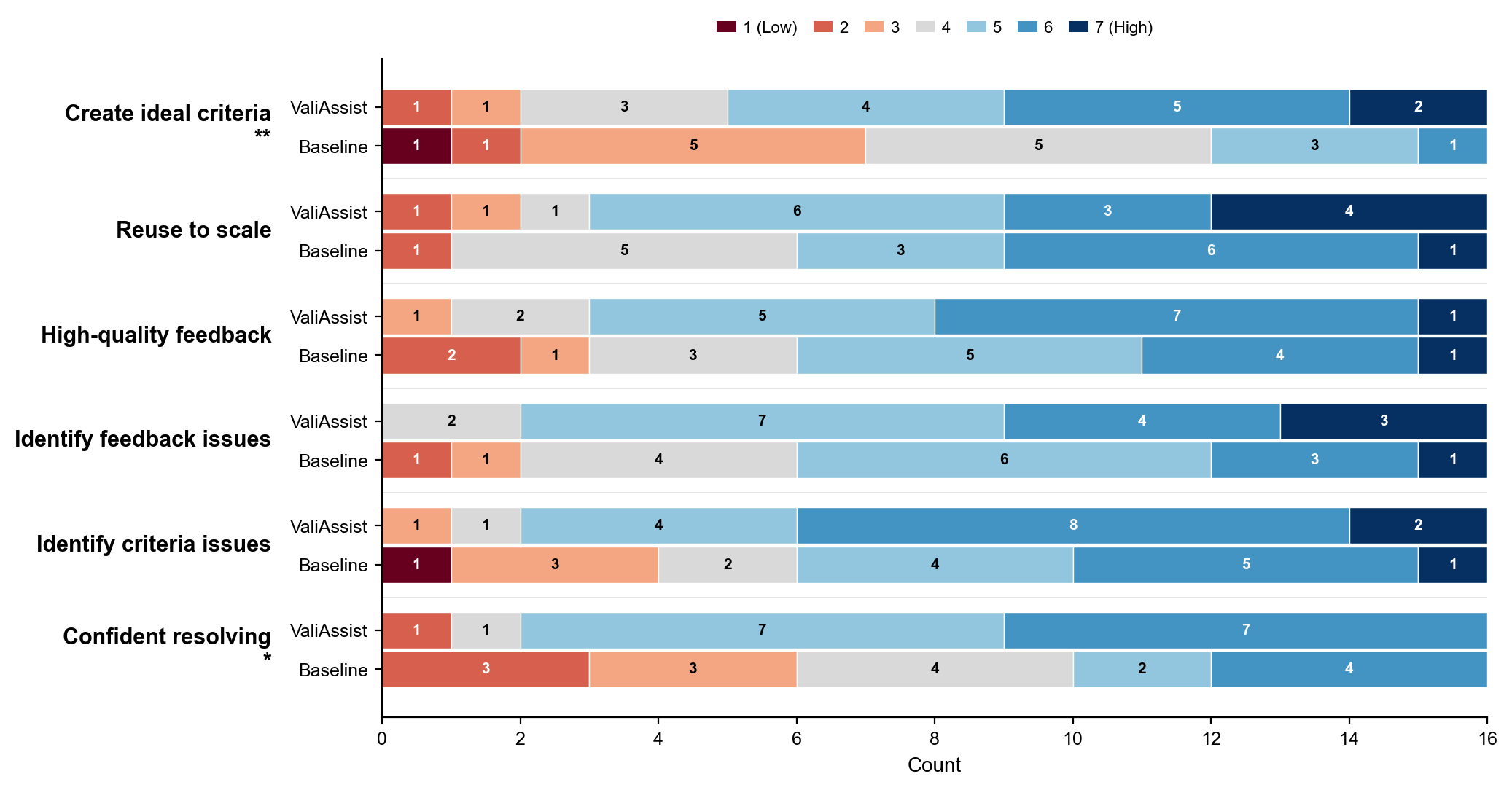}
    \caption{Participants' Agreement Level (higher is better). 1 for Strongly Disagree; 7 for Strongly Agree, * denotes $p<0.05$, and ** denotes $p<0.01$.}
    \label{fig:criteria-survey}
\end{figure}
\section{Study Results}
\subsection{Data Analysis}
For qualitative data, we coded the transcripts from the semi-structured interviews using thematic analysis. For quantitative data, we analyzed the post-condition survey's responses. Meanwhile, we analyze the quantitative data for sub-tasks and participants' interaction logs. For the feedback validation task, we did not measure the general pedagogical soundness as participants were encouraged to act on their own teaching strategies. Instead, two researchers sampled 88 validated feedback instances (up to 3 per condition per participant, some participants had fewer than 3) and measured alignment as the percentage of participant-created criteria satisfied by each feedback. On a randomly selected sample containing 30 validated feedback, we observed  Krippendorff's $\alpha=0.89$ on the independently rated alignment ratios. To capture what pedagogical intent participants were trying to achieve rather than purely measuring the numbers of criteria created in sub-task 1, two researchers used the codebook from the formative study's criteria corpus (\S~\ref{sec:formative}, Table~\ref{tab:intent-codebook}). To calibrate the coding process on the new corpus, two coders independently coded 30 randomly selected samples (30.61\% of the study corpus), and achieved a Cohen's $\kappa$ of $0.95$. They then resolved disagreements through discussion and coded the rest of the corpus independently. For both alignment and intent coding, coders were blind to the system condition and \sys{}'s \texttt{WHEN-THEN} criteria were translated to natural language criteria to avoid decision bias. For the correction sub-task, two researchers annotated 4 more code-feedback pairs regarding the matching ground truth for each criterion (i.e., a total of 16 test cases per dataset). We ran the participant-refined criteria on the test sets to assess the accuracy. We ran paired t-test for parametric data, and Wilcoxon signed-rank test for Likert-scale data.
From the interaction logs, we additionally reconstructed each participant's interaction timeline for the first two sub-tasks: every logged action is placed on a shared time axis, and an inferred \textit{stage of validation} band is derived by attributing each interval between consecutive events to the category of the event that ends it.

\subsection{\textit{Validation-by-demonstration} helped instructors create ideal validation criteria}
We observed significantly higher confidence about creating ideal criteria (Fig.\ref{fig:criteria-survey}) under \sys{} condition (\sys{}: $\mu=5.06, \sigma=1.39$; Baseline: $\mu=3.69, \sigma=1.25$, $p<0.01$). Participants also created validation criteria with more unique pedagogical intents (\sys{}: $\mu=2.56, \sigma=1.31$; Baseline: $\mu= 2.00, \sigma=0.89$, $p<0.05$). Though we did not observe statistical significance on the number of created criteria between two conditions in sub-task 1 (\sys{}: $\mu=3.00, \sigma=1.83$; Baseline: $\mu= 3.13, \sigma=2.47$), 11 of 16 participants preferred creating criteria through demonstration over description. They stated that demonstration is \textit{``less cognitively demanding''} (P2, P5, P12), and also gave them \textit{``better understanding of the system''} (P8, P9, P10, P16).
However, some participants (P1, P11) also found the demonstration limited compared to natural language descriptions. P1 expressed their worries about being \textit{``incentivized to work in individual cases''} when using demonstration in sub-task 1, before seeing their criteria applied to many feedback. This highlights the need captured by our DG3, where \sys{} in response provides both inline (Fig.\ref{fig:execution-review}B) and overview (Fig.\ref{fig:criteria-UI}G) evidence-based supports for criteria generalization.

\begin{figure}[t]
    \centering
    \includegraphics[width=\linewidth]{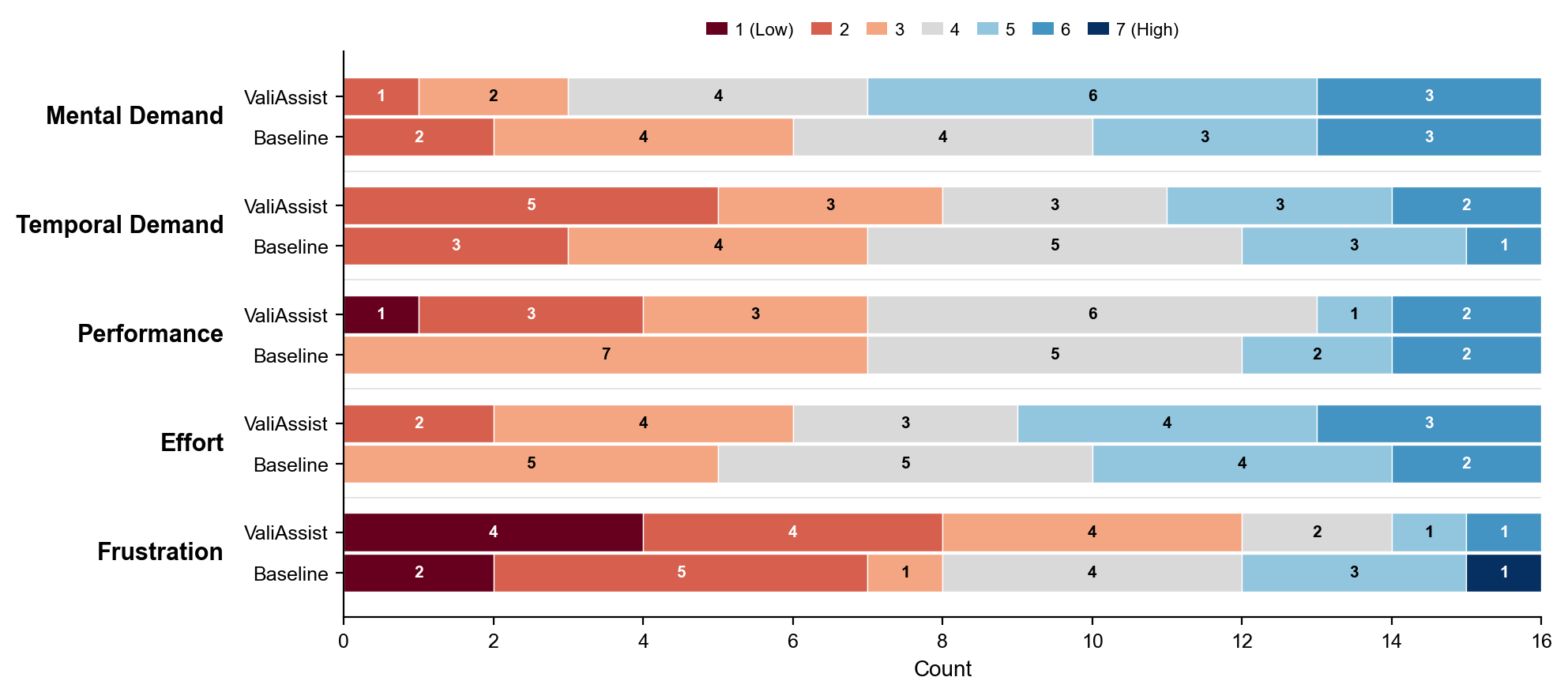}
    \caption{NASA-TLX Response. Lower is better.}
    \label{fig:nasa-tlx}
\end{figure}

\subsection{\sys{} supported more active criteria maintenance and better aligned feedback }

In sub-task 2, \sys{} enabled instructors to maintain their criteria more actively: participants conducted significantly more criterion edits---creating new criteria, paraphrasing existing ones, and adding examples---under \sys{} than Baseline (\sys{}: $\mu=2.75, \sigma=2.38$; Baseline: $\mu=0.44, \sigma=0.73$, $p<0.001$). This active maintenance translated directly into validation quality: feedback validated under \sys{} was significantly better aligned with participants' own criteria (\sys{}: $\mu=0.91, \sigma=0.12$; Baseline: $\mu=0.57, \sigma=0.34$, $p<0.01$), despite no significant difference in the raw number of feedback instances validated (\sys{}: $\mu=9.31, \sigma=5.97$; Baseline: $\mu=7.69, \sigma=8.93$) or in perceived effort and demand (Fig.\ref{fig:nasa-tlx}). Together, these results indicate that \sys{} improved validation \textit{quality} rather than sheer throughput, with no significant difference in cognitive cost. 

This quality gain appears to stem from \sys{} keeping instructors genuinely engaged in the validation loop rather than deferring to the model. Several participants (P2, P4, P6, P12, P15) noted that the demonstration approach blended naturally into their existing review workflow; P6 described it as \textit{``a more natural workflow, like, if some feedback has flaws, you will need to correct it before sending it.''} The contrast was sharpest with P2: under Baseline, P2 created 5 criteria in a ``vibing'' way, never revisited them, and continuously accepted revisions even when the feedback violated most of those criteria, a pattern of passive over-reliance on the model. Under \sys{}, P2 switched to a hands-on mode, manually refining minor edits and staying attentive to misalignments, describing demonstration as \textit{``make LLM think what I'm trying to think.''}  This shift from passive acceptance to active co-validation reflects a broader pattern in our data: \sys{}'s structured, demonstration-grounded criteria gave instructors sufficient transparency and control to treat AI assistance as a collaborator rather than an oracle.

\begin{figure}[t]
    \centering
    \includegraphics[width=1.0\linewidth]{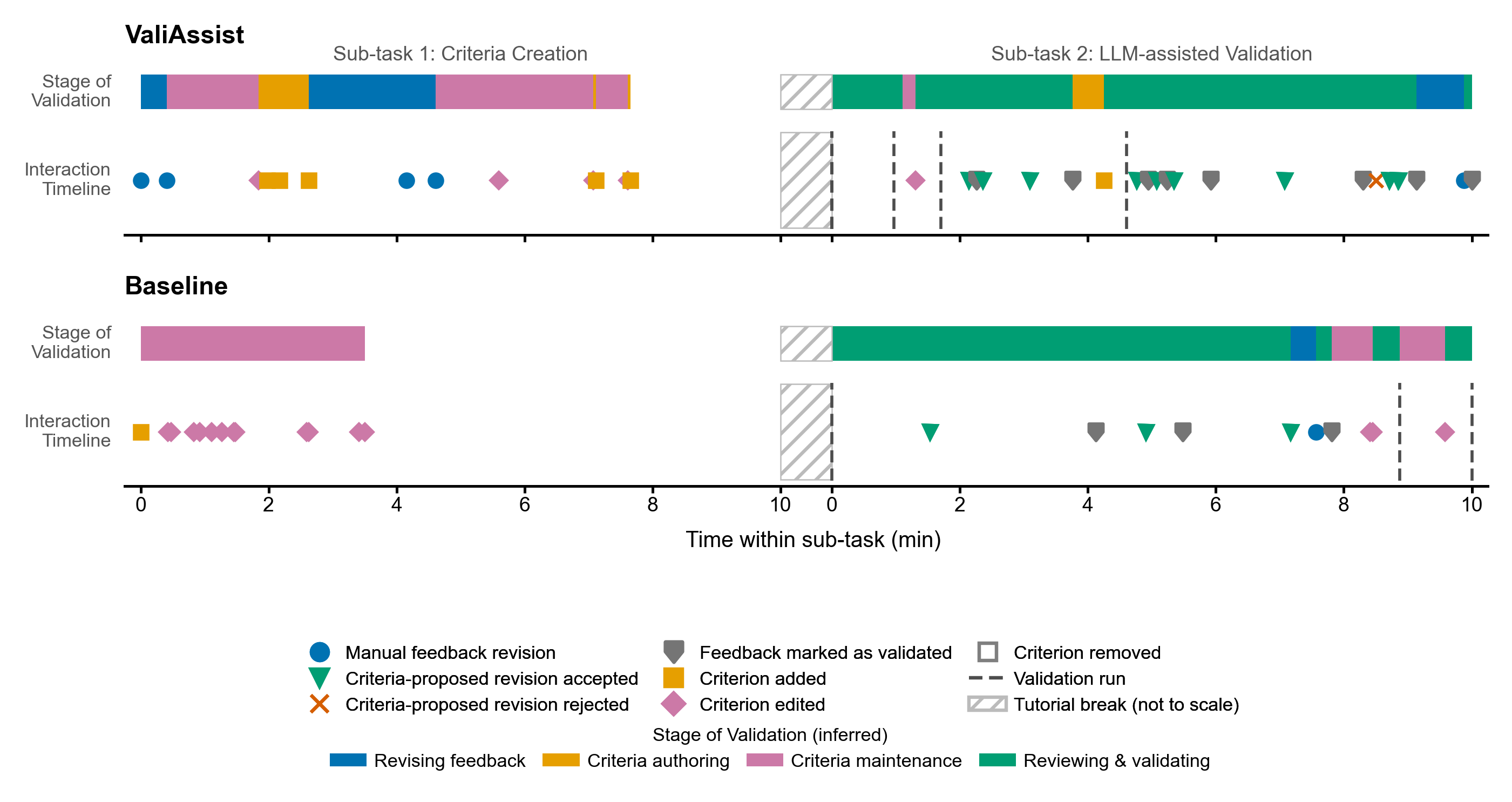}
    \caption{P12's interaction timelines during Criteria Creation (sub-task 1) and LLM-assisted Validation (sub-task 2). The top band of each panel shows the inferred stage of validation; the bottom row shows every logged action, with dashed lines marking validation runs and the hatched divider marking the untimed tutorial break (not to scale). Under \sys{}, P12 created criteria from demonstrations and short description edits, then scaled validation through five run--review cycles, accepting criteria-proposed revisions in bursts and editing criteria between runs. Under Baseline, P12 authored criteria in a single batch, spent sub-task 1 refining their wording, and applied them in fewer, later runs. Runs without adjacent markers (e.g., at the opening of sub-task 2) are re-executions performed while P12 was still reviewing the propagated results: only completed actions were logged, so reading and scrolling leave no marks.}
    \label{fig:interaction-timeline}
\end{figure}

\subsection{Instructors scaled validation through iterative criteria application}
Answering SRQ3, the interaction logs show that scaling was criteria-mediated under both conditions, but the two conditions scaled in different rhythms. In both conditions, manually revising feedback was rare (on average fewer than two revisions per participant in sub-task 2 among logged sessions); instead, participants scaled by executing criteria and reviewing the revisions those executions proposed. Under \sys{}, this reviewing was iterative: participants re-ran validation nearly twice as often as under Baseline (49 vs.\ 26 runs), reviewed smaller batches of criteria-proposed revisions per run (on average, about 6 vs.\ 10), and interleaved the criterion edits reported above between runs, so that each run-review cycle propagated their latest intent across the remaining feedback. Under Baseline, the same volume of criteria-proposed revisions instead arrived in fewer, larger sweeps.

We visualize P12's interaction timeline as an example of this contrast (Fig.\ref{fig:interaction-timeline}). Under \sys{}, P12 ran validation five times, and each run was followed by a burst of criteria-proposed revisions (18 accepted, 2 rejected) and newly validated feedback, with criterion edits between runs; under Baseline, P12 spent sub-task 1 repeatedly rewording a single batch of natural-language criteria and applied them in three later runs, accepting only 6 propagated revisions. This iterative reapplication also speaks to P1's earlier concern that demonstrations are grounded in individual cases: criteria created from single demonstrations in sub-task 1 were repeatedly reapplied to the full dataset in sub-task 2, shifting instructors' effort from revising individual feedback to supervising and refining criteria.

\subsection{\sys{} helped instructors better understand and improve the criteria}
In sub-task 3 for issue correction, participants using \sys{} achieved significantly higher accuracy (\sys{}: $\mu=0.77, \sigma=0.15$; Baseline: $\mu=0.32, \sigma=0.25$, $p<0.0001$). From the post-condition survey, participants also expressed significantly higher confidence in resolving criteria's issues (\sys{}: $\mu=5.19, \sigma=1.05$; Baseline: $\mu=4.06, \sigma=1.48$, $p<0.05$). In the interview, 13 out of 16 participants appreciated \sys{}'s structured criteria, as it was \textit{``easier to decompose ... if it's a natural language, it's hard''} (P12), \textit{`` clear and crisp''} (P8), and helped them by \textit{``keeps you some guidance about how to evaluate the AI-generated responses''} (P4). In contrast, participants mentioned the natural language criteria as \textit{``a lot more guesswork when it came to troubleshooting''} (P1) and \textit{``very hard to know what went wrong.''} (P12) Just like how P15 stated: \textit{``It's really hard for this system to point out the problem it has...I need to process a lot of aspects to evaluate my wording to say which part the system doesn't understand.''}

\section{Discussion}

\subsection{Validation by Demonstration as Progressive Specification}
Our findings suggest that validation by demonstration should not be understood as a one-shot translation from an instructor edit into a reusable rule. Instructors repeatedly applied inferred criteria, inspected their behavior on new submissions, and revised them before applying them again. In this sense, a demonstration served as a starting point for specification rather than a complete specification itself. This extends a conventional programming-by-demonstration framing in an important way. When the desired behavior depends on heterogeneous contextual evidence, the central challenge is not only inferring a program from an example, but helping users discover the boundaries of that program through its execution. For AI-assisted end-user programming systems, this suggests coupling local demonstrations with corpus-level inspection and repair: systems should expose where an inferred abstraction applies, surface counterexamples or unexpected matches, and allow users to progressively refine its scope.

\subsection{Guardrails in LLM-assisted Validation}
Our findings suggest that the key risk in LLM-assisted validation is not just inefficiency, but \textit{miscalibrated trust}. In the baseline condition, some participants treated natural-language criteria and LLM judgments as reliable enough after a first pass, even when the validated feedback still violated their own standards: One participant (P2) continued validating in a largely ``vibing'' manner despite substantial misalignment, while another (P9) achieved very low validation accuracy yet still expressed strong faith in the approach, noting that \textit{``it would be very powerful if it could work.''} This gap between actual performance and perceived promise is especially concerning in large-scale educational settings, where over-trust can allow low-quality or pedagogically misaligned feedback to propagate broadly. Our results suggest that \sys{} offers two complementary guardrails against this risk: structured criteria serve as an \textit{explicit guardrail} by making validation expectations inspectable and revisable, while validation-by-demonstration serves as an \textit{implicit guardrail} by grounding criterion creation in instructors' concrete revision work rather than abstract prompt authoring. Together, these mechanisms help keep instructors engaged in the validation process and support \textit{calibrated trust}, so that LLM assistance scaffolds judgment rather than quietly becoming unchecked delegation.

\subsection{Adaptive Criteria Creation Support Based on Context and Mental Model}
Although our lab study showed that \sys{} significantly improved criteria creation, correction, and maintenance, 5 participants (P1, P5, P9, P11, P14) still preferred using pure natural-language descriptions for criteria creation in sub-task 1 as it gave them greater expressive freedom. P9 described their ideal workflow as ``description along with demonstration.'' This preference also echoes our formative study: participants with more experience giving programming feedback were less likely to start with demonstration, as they could often articulate a concrete criterion directly. Together, these findings suggest that criteria creation support should adapt to both task context and instructors' mental models, rather than enforcing a single authoring strategy. Demonstration appears most useful when instructors are reacting to concrete failure cases or refining still-emerging intent, whereas natural-language description better supports cases where instructors already have a clear rule in mind. This suggests a broader design direction for mixed-initiative authoring: future systems should therefore support smooth transitions between description and demonstration as complementary entry points into criterion authoring. 

\subsection{Limitations}
Our lab study design has several limitations. First, though we used authentic student programming data, our evaluation was conducted in a lab setting rather than in a genuine working and grading environment. The pressures regarding misleading real students' learning progress were reduced in our setting, which could impact the attention instructors paid during feedback validation. Second, participants tested each condition of \sys{} for a limited time (i.e., about 30 minutes). Though this study could reflect instructors' behavior in short-term interactions, \sys{}'s long-term impact (e.g., reusable validation criteria across assignments) could not be examined during our lab study. Lastly, although we demonstrated \sys{}'s criteria-mediated scaling within several corpora (i.e., technical evaluation and user study), we did not run a deployment across a full term or a class of hundreds of students. Whether the same leverage holds when the criteria set grows over weeks, and when submissions arrive with error patterns the demonstrations did not cover, remains to be examined in an in-class deployment.

\section{Conclusion}

In this paper, we introduced \sys{}, a validation-by-demonstration system for scaling instructors' validation of LLM-generated programming feedback. Our study with 16 instructors showed that \sys{} supported feedback better aligned with instructors' own criteria and more accurate correction of problematic criteria than a natural-language baseline. Beyond these improvements, instructors' repeated cycles of application, inspection, and refinement suggest that demonstrations served as starting points for progressive specification. By making inferred criteria and their execution inspectable, \sys{} helped instructors clarify not only how feedback should change, but also when those changes should apply across student submissions. Together, these findings illustrate an approach to scaling expert judgment in which reusable logic extends the reach of individual decisions while remaining open to experts' inspection and refinement.

\section{Generative AI Usage Disclosure}
Generative AI tools were used to support the implementation of the system and artifact, polish the manuscript, and assist with quantitative analysis. We used Claude Code for system implementation assistance, ChatGPT and Claude for writing polish, and Claude Code for preparing Python scripts and Jupyter notebooks for data analysis. All AI-assisted coding and writing procedures were reviewed, modified as necessary, and conducted by the authors.

%
\bibliographystyle{ACM-Reference-Format}
\bibliography{ref}

\appendix

\section{Design Probe Information}
\label{apx:design_probe}
Fig\ref{fig:probe-criteria-dsl} shows the DSL in our design probe (\S\ref{sec:prob}). Our probe study informs the need for improving this DSL regarding: (1) Expressiveness (e.g., no substitution support; no style support), (2) Readability (e.g., predicates such as \texttt{hasSemanticContent} are not easy to understand for the end-user), (3) the \textbf{NOT} logic introduced extra cognitive cost, so we removed it and added \texttt{lacks}.
\begin{figure}[h]
  \centering
  \resizebox{0.7\linewidth}{!}{$
  \begin{aligned}
  \mathit{Rule} &::= \mathbf{when:}\;\Phi\;\;\mathbf{then:}\;\Delta \\[2pt]
  \Phi &::= \phi \mid \Phi \land \Phi \mid \Phi \lor \Phi \mid \lnot \Phi \\[2pt]
  \phi &::= p(t_1,\ldots,t_n) \\[2pt]
  \Delta &::= \delta \mid \Delta\;\mathbf{AND}\;\delta \\[2pt]
  \delta &::= a(u_1,\ldots,u_m) \\[4pt]
  p &::= \texttt{wordLessThan}\ \mid\ \texttt{hasExactContent}\ \mid\ \texttt{hasComponent}\ \mid\ \texttt{hasSemanticContent} \\[2pt]
  a &::= \texttt{delete}\ \mid\ \texttt{rewrite}\ \mid\ \texttt{append} \\[2pt]
  t &::= x \mid \texttt{code} \mid \texttt{feedback} 
  \qquad
  u ::= x 
  \end{aligned}
  $}
  \caption{\textbf{Probe System's Validation Criteria DSL.}
  Predicates are partitioned into code-based
  $\{\texttt{wordLessThan},\texttt{hasExactContent}\}$ and LLM-based
  $\{\texttt{hasComponent},\texttt{hasSemanticContent}\}$.
  For predicates, the first argument tags the target entity (\texttt{code} or \texttt{feedback}), while it is always \textit{feedback} for $\{\texttt{wordLessThan},\texttt{hasComponent}\}$. 
  Actions compose with \textbf{AND}, and $x$ ranges over strings.}
  \label{fig:probe-criteria-dsl}
\end{figure}

\section{Technical Evaluation}
\label{apd: techeval}
\begin{figure*}[t]
    \centering
    \includegraphics[width=1\linewidth]{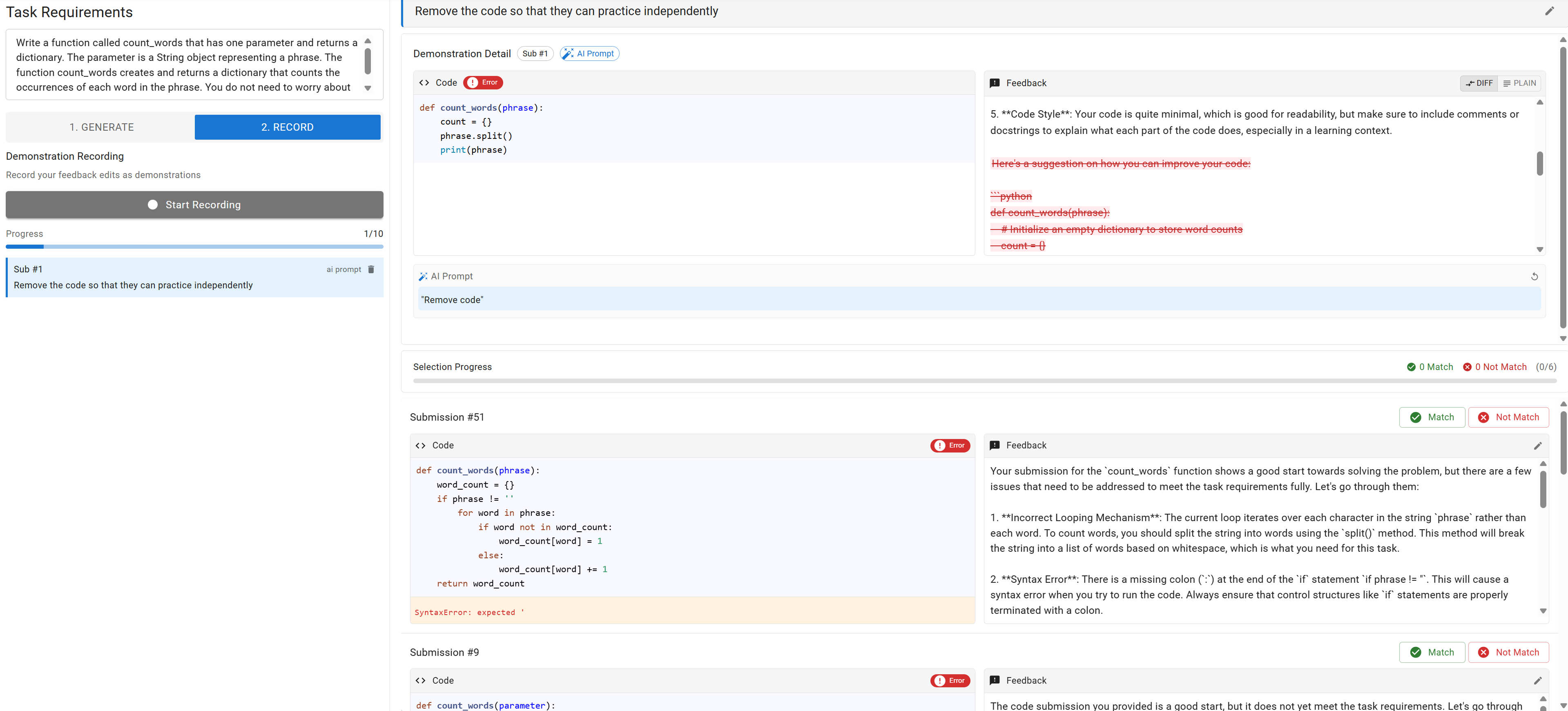}
    \caption{Revision and test case annotation UI }
    \label{fig:annotation-UI}
\end{figure*}

Fig.\ref{fig:annotation-UI} shows the UI of the data annotation used in our technical evaluation (\S\ref{sec:TechEvalDataset})

\section{Lab Study Information}
\label{apd:lab_study}
\subsection{Baseline}
\begin{figure*}[t]
    \centering
    \includegraphics[width=0.5\linewidth]{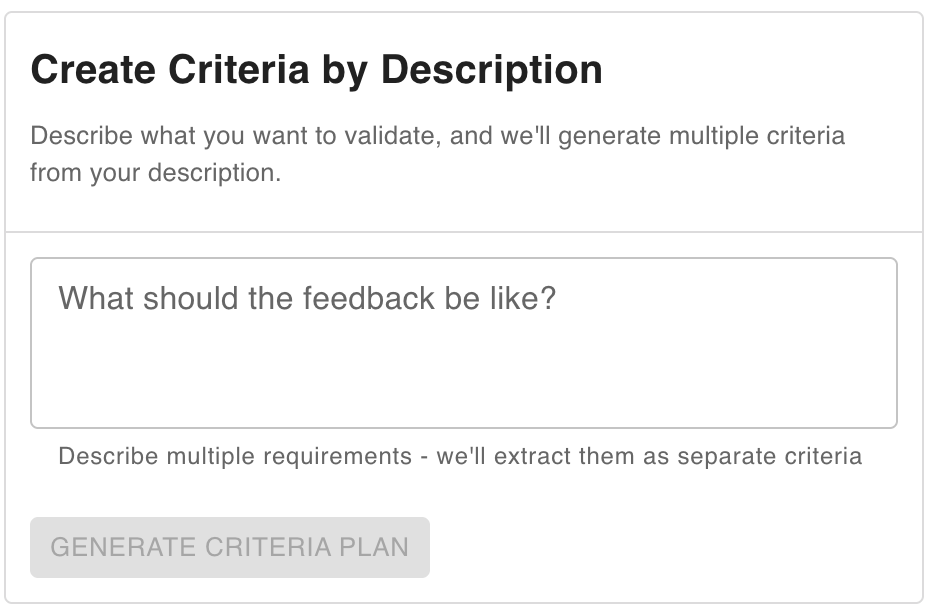}
    \caption{\textbf{Baseline in the user study.} User creates validation criteria in pure natural language with an AI assistant. They can write their criteria in prose, and the system will transform it in formal criteria. The user can refine the criteria through a multi-turn clarification.}
    \label{fig:baseline_user_study}
\end{figure*}

Fig.\ref{fig:baseline_user_study} shows the UI of the Baseline system in the user study.
\subsection{Post-condition Survey}
In our post-condition survey, we include the NASA-TLX questionnaire (without physical demand), along with an agreement collection question set of the following statements (1-7; 1 for Strongly Disagree and 7 for Strongly Agree):
\begin{itemize}
    \item ``I was able to create ideal validation criteria using the system.''
    \item ``I was able to reuse the criteria I created to scale the validation process.''
    \item ``I was able to reuse the criteria I created to produce high-quality feedback.''
    \item ``I was able to identify important issues in feedback with created criteria.''
    \item ``I was able to identify important issues in criteria by looking into the execution results.''
    \item ``I am confident that I can resolve the issues identified in the criteria.''
\end{itemize}

\end{document}